\documentclass[twocolumn]{aastex631}

\newcommand{\ourfrb}{FRB20190520B}
\usepackage{bm}
\usepackage{multirow}
\usepackage{bigdelim}
\usepackage{amsmath}

\shorttitle{AASTeX v6.31 Sample article}
\shortauthors{A. Balasubramanian et al.}
\graphicspath{{./}}

\begin{document}

\title{Continued radio observations of the persistent radio source associated with FRB20190520B provides insights into its origin}

\correspondingauthor{Arvind Balasubramanian}
\email{arvind.balasubramanian@iiap.res.in}

\author[0000-0003-0477-7645]{Arvind~Balasubramanian}
\affiliation{Indian Institute of Astrophysics, Koramangala II Block, Bangalore 560034, India}
\affiliation{Department of Astronomy and Astrophysics, Tata Institute of Fundamental Research, Mumbai, 400005, India}

\author[0000-0002-3615-3514]{Mohit~Bhardwaj}
\affiliation{McWilliams Center for Cosmology, Department of Physics, Carnegie Mellon University, Pittsburgh, PA 15213, USA}

\author[0000-0003-2548-2926]{Shriharsh~P.~Tendulkar}
\affiliation{Department of Astronomy and Astrophysics, Tata Institute of Fundamental Research, Mumbai, 400005, India}
\affiliation{National Centre for Radio Astrophysics, Post Bag 3, Ganeshkhind, Pune, 411007, India}
\affiliation{CIFAR Azrieli Global Scholars Program, MaRS Centre, West Tower, 661 University Ave, Suite 505, Toronto, ON, M5G 1M1 Canada}
\begin{abstract}
Follow-up studies of persistent emission from Fast Radio Burst (FRB) sources are critical for understanding their elusive emission mechanisms and the nature of their progenitors. This work presents new observations of the persistent radio source (PRS) associated with \ourfrb. We observe a gradual decay in the PRS brightness, which is punctuated by periods of brightening and dimming at both 1.5\,GHz and 3\,GHz. Furthermore, our low-frequency ($<1$\,GHz) observations—the first for this source—reveal evidence of a spectral break which can be attributed to absorption processes. Interpreted within the framework of the magnetar wind nebula model, our data constrain the age of the magnetar progenitor to $52^{+16}_{-10}$\,years, broadly consistent with previous work. Assuming the observed 1.5\,GHz variability is driven by scintillation, we discuss the constraints on the size of the persistent source. The observations presented here challenge the predictions of the previously published best-fit hypernebula model for this source.

\end{abstract}

\keywords{\href{http://astrothesaurus.org/uat/1339}{Radio bursts (1339)}, \href{http://astrothesaurus.org/uat/2008}{Radio transient sources (2008)}, \href{http://astrothesaurus.org/uat/1851}{Transient sources (1851)}, \href{http://astrothesaurus.org/uat/1358}{Radio sources (1358)}}


\section{Introduction} \label{sec:intro}

Fast Radio Bursts (FRBs) are highly energetic $\sim$\,millisecond duration radio transients \citep{2007Sci...318..777L}. The observations of FRBs using various
radio facilities have revealed that some FRBs repeat, while others appear to be non-repeating \citep{2022A&ARv..30....2P}. Among the FRBs that repeat, only two have shown evidence of a periodicity in their activity windows \citep[][and refs. therein]{2022A&ARv..30....2P}. FRBs experience propagation effects due to interaction and scattering before reaching our detectors. Some propagation effects include: dispersion measure (DM)—the delay in arrival of low frequency radiation compared to the high frequency radiation due to interaction with ionized medium; rotation measure (RM)—Faraday rotation of the polarization angle due to magnetic field along the line of sight. The measured DM of the FRB can be expressed as a sum of the DM contribution from the Milky Way, the interstellar medium, the intergalactic medium, and the local DM from the host galaxy of the FRB.

The origin of FRBs is still an open question, with many proposed models \citep{2019PhR...821....1P}. The discovery of an FRB-like burst from galactic magnetar SGR1935+2154 has bolstered interest in magnetar-based models as the leading candidates for explaining FRBs \citep{2020Natur.587...59B,2020Natur.587...54C}. To date, a total of $\approx1000$ FRB sources have been identified, of which $<10\%$ are repeaters \citep{2021ApJS..257...59C, 2024ApJ...969..145C, 2023Univ....9..330X, universe11070206}. Less than a 100 FRBs have been localized to their host galaxies, providing rich insights into FRB environments \citep[see e.g.][]{ 2017Natur.541...58C,2021ApJ...908L..12T,2020Natur.577..190M,2022Natur.606..873N}. Among the population of repeating FRBs, a handful of sources are of particular interest given that a compact persistent radio source (PRS) have been observed associated with them—FRB20121102 \citep{2017Natur.541...58C}, FRB20190520B \citep{2022Natur.606..873N}, FRB20201124A \citep{2024Natur.632.1014B, 2025ApJ...993..234B} and FRB20190417A \citep{2025arXiv250905174M} and FRB20181030A \citep{2024ApJ...976..199I}.

\ourfrb\ was discovered by the Five-hundred-meter Aperture Spherical radio Telescope (FAST) on May 20, 2019 as part of the CRAFT Survey \citep{2018IMMag..19..112L}. It has been localized to a star-forming dwarf galaxy at a redshift of $z=0.241$ \citep{2022Natur.606..873N, 2025ApJ...982..203C}. The average DM of bursts from this source is 1207\,cm$^{-3}$pc. Initially, the DM contribution from the host was thought to be exceptionally large $\sim900$\,cm$^{-3}$pc. However, \cite{2023ApJ...954L...7L} investigated the effect of foreground galaxies and estimated a substantially lower host DM. \cite[also see][]{Bhardwaj_2025}. Observations of bursts from this FRB using the Green Bank Telescope (GBT) and Parkes telescope by \citep{2023Sci...380..599A} have shown that the RM changes sign twice, pointing to a change in the magnetic field component parallel to the propagation, possibly due to a turbulent magnetized screen surrounding the FRB source. \cite{2023ApJ...958L..19B} used Very-long-baseline interferometry (VLBI) observations to constrain the size of the PRS $<9$\,pc and co-located it to within $80$\,pc of the FRB. This PRS exhibits a flat spectrum (slope of $-0.4$ at 2020 and $-0.33$ at 2021 epochs) above 1\,GHz accompanied by a $\sim20$\% decline in broadband flux between the 2020 and 2021 epochs \citep{2023ApJ...959...89Z}. The PRS has not been detected at frequencies $<1$\,GHz in the past. 

Some leading models proposed to explain persistent radio emission from FRB sources are—Magnetar Wind Nebula (MWN) model \citep{2018ApJ...868L...4M} and hypernebula model \citep{2022ApJ...937....5S}. The Magnetar Wind Nebula model attributes persistent radio emission to synchrotron emission from magnetized electron-ion interactions with the nebula, powered by flaring activity of a young magnetar. The Hypernebula model explains the persistent emission as the interaction of winds emitted by an accreting black hole with the surrounding medium. Details of the interpretations of these models are discussed in Section \ref{sec:analysis}.

In this paper, we present the first ever low frequency  observations of the persistent radio source associated with \ourfrb. We interpret the temporal and spectral evolution of the persistent emission by combining our new observations and previous observations of the source. Section \ref{sec:observations} describes the observations and data reduction performed, followed by Section \ref{sec:analysis} covering the analysis of the observations. Finally, we conclude with a discussion of our analysis in Section \ref{sec:conclusion}.

\section{Observations}
\label{sec:observations}

\subsection{uGMRT observations and archival data}

We observed the PRS associated with \ourfrb\ using the wideband receiver backend of upgraded Giant Metrewave Radio Telescope (uGMRT) in three frequency bands:  band 3 (central frequency, $\nu_{c}$ = 400\,MHz, bandwidth, BW = 200\,MHz), band 4 ($\nu_{c}$ = 750\, MHz, BW =  400\,MHz) and band 5 ($\nu_{c}$ = 1260\,MHz, BW = 400\,MHz) between 2023 June 16 and 2023 June 20 (Proposal 44\_039, PI: Balasubramanian). The band 3 observations under this project were split over two days. 
We also obtained three epochs of observations each in band 4 and band 5 between 2024 September 05 and 2024 September 21 (Proposal 46\_126, PI: Balasubramanian). In addition to our observations, we included archival uGMRT data from proposal 43\_054 (PI: Yi Feng) in our analysis. Raw data were downloaded in the \texttt{FITS} format and converted to the \texttt{CASA} \citep{2022PASP..134k4501C} measurement set format. The data were then calibrated and imaged using the automated continuum imaging pipeline \texttt{CASA-CAPTURE} \citep{2021ExA....51...95K}. All observations used 3C286 as the flux calibrator. J1543-0757 was used as the phase calibrator for observations under 43\_054 and 46\_126 while J1558-1409 was used as the phase calibrator for observations under 44\_039. For the band 3 data, the two measurement sets (under 44\_039) were passed onto the automated pipeline separately and then imaged after combining the calibrated measurement sets. Each pipeline run included eight rounds of self-calibration.  
The band 5 flux density of the persistent source was calculated using the \texttt{CASA} task \texttt{imfit} within a small circular region (of radius $\sim2\times$ the size of the synthesized beam at band 5) centered at the PRS coordinates. An additional 5\% flux density error was added in quadrature to the error obtained from \texttt{imfit} to account for flux calibration errors (see Table \ref{tab:comprehensive_tab}). No detections were made in Band 3 or in any epoch of Band 4 observations. The upper limit values listed in Table \ref{tab:comprehensive_tab} are the $3\times$RMS value within a large circular region (of radius $\sim20\times$ the size of the synthesized beam at the respective band) centered at the position of the PRS. None of the upper limits are used in the analyses in the section that follows. The observation on 2022 Nov 11 (under GMRT 43\_053) shows a possible phase calibration issue, which was identified by imaging the flux calibrator using the same calibration solution (see Figure \ref{fig:upper_limits} in appendix for a comparison of the non-detection versus detection of the PRS and the flux calibrator). Additionally, the two observations on 2024 Sep 15 and 2024 Sep 21 (under 46\_126) were affected by severe radio frequency interference, not allowing for sufficiently clean images.

\subsection{VLA observations and archival data}
We conducted a single L-band ($\sim$1.5\,GHz) observation of the PRS using the Jansky Very Large Array (VLA) on 2024 August 08. In addition,  VLA S-band ($\sim$3\,GHz) data of the PRS20190520B field were obtained from the archival data listed under proposal 23A-010 (PI: Yi Feng, taken between 2023 June 07 and 2023 June 24). We obtained the calibrated measurement set when available or started with the raw data. Raw data were calibrated using the automated VLA calibration pipeline \footnote{\url{https://science.nrao.edu/facilities/vla/data-processing/pipeline/CIPL_654}}. All observations used 3C286 as the flux calibrator. J1543-0757 was used as the phase calibrator for observations under 23A-010, while J1558-1409 was used as the phase calibrator for observations under 24A-409. Following calibration, the data were imaged using the automated imaging pipeline \footnote{\url{https://science.nrao.edu/facilities/vla/data-processing/pipeline/vipl_666}} and refined through self-calibration. The best self-calibrated image was selected, and the \texttt{CASA} task \texttt{imfit} was used in a small region around the PRS to estimate the source flux density. An additional 5\% flux uncertainty was added in quadrature to account for flux calibration errors. It may be noted that the data under 23A-010 has already been analyzed using custom software based on the Astronomical Image Processing System\footnote{\url{https://www.aips.nrao.edu/index.shtml}} by \cite{2024ApJ...976..165Y}. Here, we repeated the analysis of this data using the CASA VLA calibration pipeline to maintain consistency with the other measurements in this work. There are slight differences in the obtained measurements compared to the ones discussed in \cite{2024ApJ...976..165Y} due to the different calibration and imaging tools used here.

A summary of all measurements is listed in Table \ref{tab:comprehensive_tab}. Figure \ref{fig:mann_kendall} shows the decay of the overall flux density of the PRS over time. Figure \ref{fig:190520_lc_joint_fit} shows the temporal variation of the PRS flux density at different frequency bands, and Figure \ref{fig:190520_spec_joint_fit} displays the variation of the spectrum over the epochs. We observe flaring and dimming episodes of the source in the $\approx1.5$\,GHz data, especially in the late 2020 epoch. The possible reason for this observation is discussed in the Section \ref{subsec:var_scintil}. For some observations, the image was heavily affected by radio frequency interference (RFI), resulting in an image that was insufficiently clean to use for our analysis (see Table \ref{tab:comprehensive_tab} footnote). 

\startlongtable\begin{deluxetable*}{cccccc}
\tablecaption{Summary of radio continuum observations of the persistent radio source associated with \ourfrb. For non-detections, 3\,$\sigma$ upper limits are reported, where $\sigma$ is the RMS flux density measured in a large region of the residual image. Observation dates (in two formats),  frequency bands, flux densities and their associated errors, and proposal identifiers/paper references are listed.
\label{tab:comprehensive_tab}}
\tablecolumns{6}
\tablehead{\colhead{Date} & \colhead{Time (MJD)} & \colhead{Frequency (GHz)} & \colhead{Flux density (uJy)} & \colhead{Flux density error (uJy)} & \colhead{Reference}} 
\startdata
2020 Jul 21 & 59051.067 & 1.5 & 258 & 29 & \cite{2023ApJ...959...89Z} \\
2020 Jul 23 & 59053.08 & 1.5 & 273 & 37 &  \\
2020 Aug 18 & 59079.019 & 5.5 & 145 & 17 &  \\
2020 Aug 18 & 59079.986 & 5.5 & 164 & 19 & \\
2020 Aug 29 & 59090.956 & 5.5 & 158 & 17 &  \\
2020 Aug 30 & 59091.953 & 3.0 & 195 & 24 &  \\
2020 Sep 12 & 59104.881 & 3.0 & 160 & 21 &  \\
2020 Sep 12 & 59104.923 & 5.5 & 151 & 17 &  \\
2020 Sep 13 & 59105.991 & 3.0 & 186 & 24 &  \\
2020 Sep 15 & 59107.93 & 5.5 & 153 & 17 &  \\
2020 Sep 19 & 59111.13 & 3.0 & 176 & 25 &  \\
2020 Nov 08 & 59161.691 & 5.5 & 139 & 20 &  \\
2020 Nov 14 & 59167.652 & 3.0 & 233 & 29 &  \\
2020 Nov 16 & 59169.655 & 3.0 & 211 & 25 &  \\
2021 Oct 01 & 59488.879 & 10.0 & 115 & 24 & \cite{2023ApJ...959...89Z} \\
2021 Oct 01 & 59488.883 & 5.5 & 114 & 28 &  \\
2021 Oct 01 & 59488.887 & 3.0 & 112 & 34 &  \\
2021 Oct 01 & 59488.895 & 1.5 & 240 & 70 &  \\
2021 Nov 07 & 59525.861 & 10.0 & 81 & 18 &  \\
2021 Nov 07 & 59525.865 & 5.5 & 139 & 33 &  \\
2021 Nov 07 & 59525.869 & 3.0 & 111 & 33 &  \\
2021 Nov 07 & 59525.877 & 1.5 & 212 & 61 &  \\
\tableline
2022 Feb 26 & 59636.208 & 1.7 & 197 & 34 & \cite{2023ApJ...958L..19B} \\
2022 Feb 27 & 59637.208 & 1.7 & 210 & 34 &  \\
\tableline
\tablenotemark{$\dagger$} 2022 Nov 11 & 59894.205 & 1.3 & \nodata & \nodata & GMRT 43\_054 (This work)\\
2022 Nov 29 & 59912.125 & 1.3 & 268 & 44 &  \\
2022 Dec 27 & 59940.131 & 1.3 & 156 & 26 &  \\
2023 Jan 24 & 59968.036 & 1.3 & 281 & 41 &  \\
2023 Feb 21 & 59996.167 & 1.3 & 119 & 36 &  \\
2023 Mar 21 & 60024.962 & 1.3 & 280 & 25 &  \\
\tableline
2023 Jun 07 & 60102.323 & 3.0 & 157 & 9 & VLA 23A-010 (This work) \\
2023 Jun 15 & 60110.234 & 3.0 & 173 & 11 &  \\
\tableline
2023 Jun 16 & 60111.84 & 0.7 & $<$165 & \nodata & GMRT 44\_039 (This work) \\
\tableline
2023 Jun 17 & 60112.192 & 3.0 & 171 & 12 & VLA 23A-010 (This work) \\
2023 Jun 18 & 60113.178 & 3.0 & 166 & 10 & \\
2023 Jun 18 & 60113.219 & 3.0 & 165 & 11 & \\
\tableline
2023 Jun 19 & 60114.196 & 0.3 & $<$195 & \nodata & GMRT 44\_039 (This work) \\
\tableline
2023 Jun 20 & 60115.172 & 3.0 & 152 & 10 & VLA 23A-010 (This work) \\
2023 Jun 20 & 60115.214 & 3.0 & 151 & 11 &  \\
2023 Jun 20 & 60115.255 & 3.0 & 143 & 10 &  \\
2023 Jun 20 & 60115.297 & 3.0 & 152 & 10 &  \\
\tableline
2023 Jun 20 & 60115.584 & 1.3 & 179 & 18 & GMRT 44\_039 (This work) \\
\tableline
2023 Jun 23 & 60118.262 & 3.0 & 147 & 13 & VLA 23A-010 (This work) \\
2023 Jun 24 & 60119.168 & 3.0 & 155 & 15 &  \\
2023 Jun 24 & 60119.209 & 3.0 & 144 & 13 &  \\
\tableline
2024 Aug 08 & 60530.985 & 1.5 & 221 & 13 & VLA 24A-409 (This work) \\
\tableline
2024 Sep 03 & 60556.588 & 1.3 & 230 & 24 & GMRT 46\_126 (This work) \\
2024 Sep 05 & 60558.585 & 0.7 & $<$155 & \nodata &  \\
2024 Sep 12 & 60565.589 & 1.3 & 184 & 16 &  \\
\tablenotemark{$\dagger\dagger$}  2024 Sep 15 & 60568.594 & 0.7 & \nodata & \nodata &  \\
2024 Sep 21 & 60574.329 & 1.3 & 190 & 20 &  \\
\tablenotemark{$\dagger\dagger$}  2024 Sep 21 & 60574.0 & 0.7 & \nodata & \nodata &  \\
\enddata
\tablenotetext{\dagger}{Probable issue with phase calibration. See non-detection in Figure \ref{fig:upper_limits}.}
\tablenotetext{\dagger\dagger}{ Data affected by severe radio frequency interference. Sufficiently clean image was not obtained.}
\end{deluxetable*}

\section{Physical Interpretations}
\label{sec:analysis}

Equipped, with the newly observed multi-frequency dataset of the PRS associated with \ourfrb, we present constraints on the physical properties of the possible progenitor of the persistent emission.

\subsection{Monotonic trend in the observations}
\begin{figure}[h]
    \centering
    \includegraphics[width=\columnwidth]{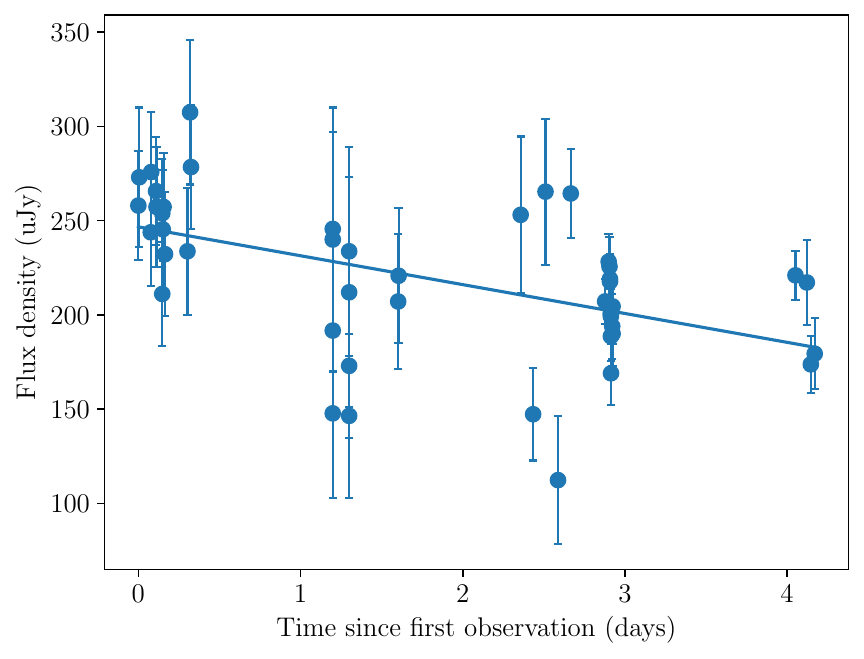}
    \caption{Light curve with all the flux density measurements from Table \ref{tab:comprehensive_tab} scaled to 1.5\,GHz assuming a spectral index of -0.4. The fit is a linear regression showing a decreasing trend in the overall flux density over time.}
    \label{fig:mann_kendall}
\end{figure}

To investigate presence of a monotonic trend in the light curve, we took all the observations listed in Table \ref{tab:comprehensive_tab}, and scaled the flux densities to 1.5\,GHz assuming a spectral index of $-0.4$ \citep[as observed in ][ for the 2020 epoch]{2023ApJ...959...89Z}. This scaled data was then subjected to the Mann–Kendall test using the \texttt{pymannkendall} package \citep{Hussain2019pyMannKendall}. The test yielded a p-value of $1.9\times10^{-5}$ and a Kendall Tau value of $-0.4$, suggesting \textbf{a significant decreasing trend}. Figure \ref{fig:mann_kendall} shows the light curve with the scaled flux density values along with the linear regression fit (slope $=-15.3$) performed with \texttt{scipy} package \citep{2020SciPy-NMeth} showing the decreasing trend. Changing the assumed spectral index to $-0.33$ \citep[as observed in ][ for the 2021 epoch]{2023ApJ...959...89Z} results in similar Mann Kendall p-value ($2.8\times10^{-4}$), Kenadall Tau value ($-0.4$) and linear regression fit (slope $=-11.8$) as above. This establishes a decreasing trend in the total flux density of the PRS as a function of time.

\subsection{Magnetar Wind Nebula Model}\label{subsec:PWN}
\cite{2016MNRAS.461.1498M} and \cite{2018ApJ...868L...4M} have shown that a single expanding magnetized electron-ion nebula powered by a young magnetar can interact with the surrounding medium to produce synchrotron emission. The magnetar is assumed to inject energy into the nebula at a rate given by 

\begin{equation}
    \dot{E} = (\alpha-1)\frac{E_{B_{\star}}}{t_{0}} \left( \frac{t}{t_{0}}\right)^{-\alpha} \hspace{5pt} \rm{for} \hspace{5pt} t\geqslant t_{0}, \hspace{5pt} \alpha > 1
\end{equation}

where $E_{B_{\star}}$ is the free magnetic energy of the magnetar, $\alpha$ is a powerlaw index, and $t_{0}$ is the onset of the active period of the magnetar. The energy injection drives electrons through the surrounding medium, producing the observed persistent emission.
At frequencies $\nu$ above the characteristic synchrotron self-absorption frequency $\nu_{SSA}$, the synchrotron luminosity decays as follows \citep[see][for detailed calculations]{2018ApJ...868L...4M} 

\begin{equation}
\label{eq:L_nu}
    L_{\nu} \propto \nu^{-\left(\frac{\alpha-1}{2}\right)} t^{-\frac{\alpha^{2} + 7\alpha - 2}{4}}     
\end{equation}

\begin{figure}[h]
    \centering
    \includegraphics[width=\columnwidth]{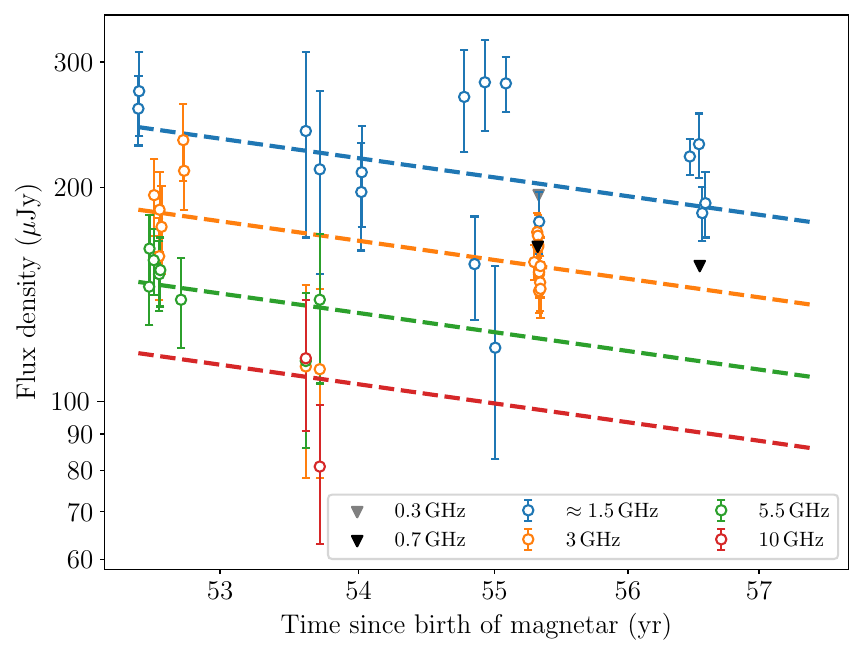}
    \caption{Light curve of the \ourfrb\ PRS, showing the joint spectro-temporal fit using Equation \ref{eq:joint_fit}. This fit gives an estimate of the age of the persistent source to be $t_{\rm{age}}=52$\,years.}
    \label{fig:190520_lc_joint_fit}
\end{figure}

\begin{figure}[h]
    \centering
    \includegraphics[width=\columnwidth]{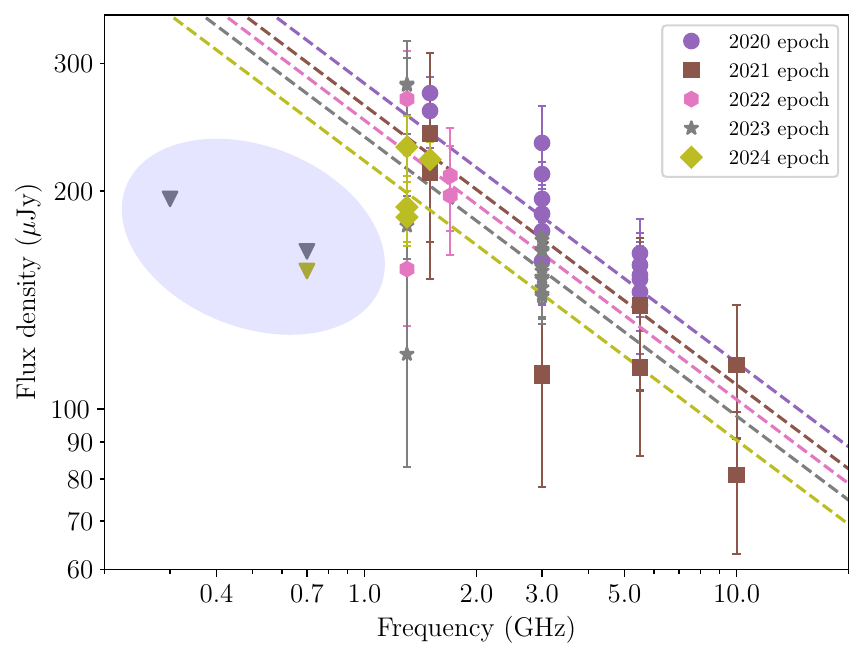}
    \caption{Spectrum of the PRS of \ourfrb\ showing the joint spectro-temporal fit using Equation \ref{eq:joint_fit}. The color of the dashed lines represent the different epochs. The shaded blue region shows the uGMRT upper limits that suggest a break in the spectrum at frequencies $<1$\,GHz.}
    \label{fig:190520_spec_joint_fit}
\end{figure}

We adopt Equation \ref{eq:L_nu} and modify it as follows  
\begin{equation}\label{eq:joint_fit}
    F(\nu,t) = A \, \nu^{-\left(\frac{\alpha-1}{2}\right)} \left(\frac{t_{\rm{obs}}+t_{\rm{age}}}{t_{\rm{age}}}\right)^{-\left(\frac{\alpha^{2} +7\alpha - 2}{4}\right)},
\end{equation}

where $A$ is a scaling constant, $t_{\rm{obs}}$ is the time since the first observation of the PRS, $t_{\rm{age}}$ is the age of the persistent source at the time of the first observation. We perform an MCMC fit of Equation \ref{eq:joint_fit} using the \texttt{emcee} package \citep{2013PASP..125..306F}, incorporating all the detections listed in Table \ref{tab:comprehensive_tab}. The allowed range of the parameters are $\alpha>1$ (from the assumption of the MWN model), $4$\,yr $<t_{\rm{age}}<1900$\,yr \citep[from][]{2023ApJ...958L..19B} and $A>1$\,$\mu$Jy (a scaling paramter that is greater than 1\,$\mu$Jy can be deduced visually from Figure \ref{fig:mann_kendall}). The posterior distributions were analyzed using \texttt{ChainConsumer} \citep{Hinton2016} resulting in the best-fit parameters listed in Table \ref{tab:fit_results}. Figure \ref{fig:190520_lc_joint_fit} shows the best fit light curves for different frequencies, following the color-code of the data points. The corresponding best fit spectra are shown in Figure \ref{fig:190520_spec_joint_fit}. The best-fit age of the persistent source, $t_{\rm{age}}$, is $52^{+16}_{-10}$\,years, consistent with the allowed age limits derived in \cite{2023ApJ...958L..19B}, slightly higher than the 16--22\,year estimate from \cite{2021ApJ...923L..17Z} and more closer to the estimate of 40\,years from the magnetar-flare powered model discussed in \cite{2024arXiv241219358B}. The best-fit energy injection power law index is $\alpha=1.77^{+0.07}_{-0.08}$. Excluding the flaring and dimming events observed in late 2022 does not significantly alter the best-fit parameter values.

\begin{deluxetable}{|c|c|c|c|}
\label{tab:fit_results}
\tablecaption{MCMC posterior best fit parameters for the decay using Equation \ref{eq:joint_fit}.}
\tablehead{Parameter & Prior type & Bounds & Posterior value}
\startdata
$\alpha$  & uniform & (1, $\infty$) & $1.77^{+0.07}_{-0.08}$\\
$t_{\rm{age}}$ (yr)  & uniform & (4,1900) &  $52^{+16}_{-10}$\\
$A$ (uJy)  & uniform & (1, $\infty$) & $284^{+18}_{-19}$\\
\enddata
\end{deluxetable}

In addition to the flux decay observed at frequencies $\nu > 1$\,GHz, the upper limits from our uGMRT observations at $\nu < 1$\,GHz (from the 2023 and 2024 epochs) suggest  a spectral break between 700\,MHz and 1.3\,GHz (see blue shaded portion in  Figure~\ref{fig:190520_spec_joint_fit}). At the observed timescales, such a spectral break is consistent with synchrotron self-absorption. However, the available upper limits are insufficient to constrain the synchrotron self-absorption frequency. Deeper observations at sub-GHz frequencies will be required to quantitatively determine the location of the spectral break. A recent study of the persistent emission from FRB20240114A using GMRT, VLA and MeerKAT data, shows evidence of synchrotron self-absorption peak at $\nu\approx1.6$\,GHz \citep{2025arXiv250114247Z}, thereby agreeing with the spectral break below $1.3$\,GHz that we observe for the PRS associated with \ourfrb\ in this work.

\subsection{Scintillation as a source of PRS variability}\label{subsec:var_scintil}
\cite{2023ApJ...959...89Z} analysed temporal variability above 1 GHz and found no significant variation in most bands. A marginal variability was seen at 3\,GHz. We extend the same variability analysis, adding our 1.3\,GHz observations to the 1.5\,GHz observations reported by \cite{2023ApJ...959...89Z}.  One of the causes of variability of a compact radio source is interstellar scintillation. Scintillation arises due to inhomogeneities in the ionized ISM and can cause apparent flux variations in compact radio sources. We follow the formalism introduced in \cite{1992RSPTA.341..151N} and further explored in \cite{1998MNRAS.294..307W} for our analysis here. The scattering strength parameter $\xi$ describes the scintillation due to a screen at a distance $D$ from the observer. $\xi=1$ corresponds to the critical value at which phase changes are substantial, across the characteristic first Fresnel zone $\theta_{F}=\sqrt{c/2\pi \nu D}$, where $\nu$ is the frequency of observation. Now, $\xi$ is expressed as $(\nu_{0}/\nu)^{17/10}$, where $\nu_{0}$ is the transition frequency between the weak scattering ($\xi\ll1$ and $\nu_0<\nu$), and strong scattering ($\xi\gg1$ and $\nu_0>\nu$) regimes. This transitional frequency has been estimated to be $\nu_{0}=12.53$\,GHz using the galactic free electron density model in pyNE2001 \citep{2002astro.ph..7156C}. Given $\nu_{0}=12.53$\,GHz, we are in the strong scattering regime as $\nu_0>\nu$ (as $\nu=1.5$\,GHz). There are two possible scintillation scenarios: diffractive scintillation and refractive scintillation. Diffractive scintillation is relevant at characteristic frequency bandwidth, of $\xi^{-2}\nu = \nu(\nu_{0}/\nu)^{17/5}$ which is $\sim1$\,MHz for our observations. Our observations span a larger bandwidth, washing out diffractive scintillation. Hence, refractive scintillation is the relevant mechanism here. For, refractive scintillation, the expected modulation index is 
\begin{equation}
m_{\rm{exp}} = \xi^{-1/3} = \left( \frac{\nu}{\nu_{0}} \right)^{17/30} = 0.29
\end{equation}

 Assuming a scattering disk at a distance of $D=1$\,kpc from us, the size of the Fresnel zone at $\nu=1.4$\,GHz is $\theta_{F}=\sqrt{c/2\pi \nu D} = 6.78\,\mu\rm{as}$. Therefore, the angular size of the screen is 
 
 \begin{equation}
 \theta_{r} = \theta_{F}\xi = 284.1\,\mu\rm{as}
 \end{equation}
 
This angular size of the screen can be converted to a projected radius of the screen at the distance of the source using the luminosity distance of the host galaxy (i.e. $d_{\rm{lum}}=1218$\,Mpc) as follows:

\begin{equation}
    R_{r} = \theta_{r}d_{\rm{lum}}/2 = 5.0\times 10^{18}\,\rm{cm} = 1.68\,\rm{pc}
\end{equation}

The characteristic timescale associated with the refractive scintillation from this screen is 

\begin{equation}
t_{r} = 2 \left( \frac{\nu_{0}}{\nu} \right)^{11/5} = 10.47\,\rm{days}
\end{equation}

Given this expected timescale, we compute the modulation index of the observed data in the frequency range 1.3--1.7\,GHz from Table \ref{tab:comprehensive_tab}, by averaging consecutive data that are within the refractive scintillation timescale, $t_r$, using the following expression:

\begin{equation}
m_{\rm{obs}} = \frac{1}{\overline{F_{i}}} \sqrt{\frac{N}{N-1} \left(\overline{F_{i}^2} - \overline{F_{i}}^2\right)}
\end{equation}

where $F_{i}$ are the flux density measurements, $\overline{F_{i}}$ is the mean of the flux density measurements, $\overline{F_{i}^2}$ is the mean of the squares of $F_{i}$ and $N$ is the total number of observations.

\begin{figure}[h]
    \centering
    \includegraphics[width=\columnwidth]{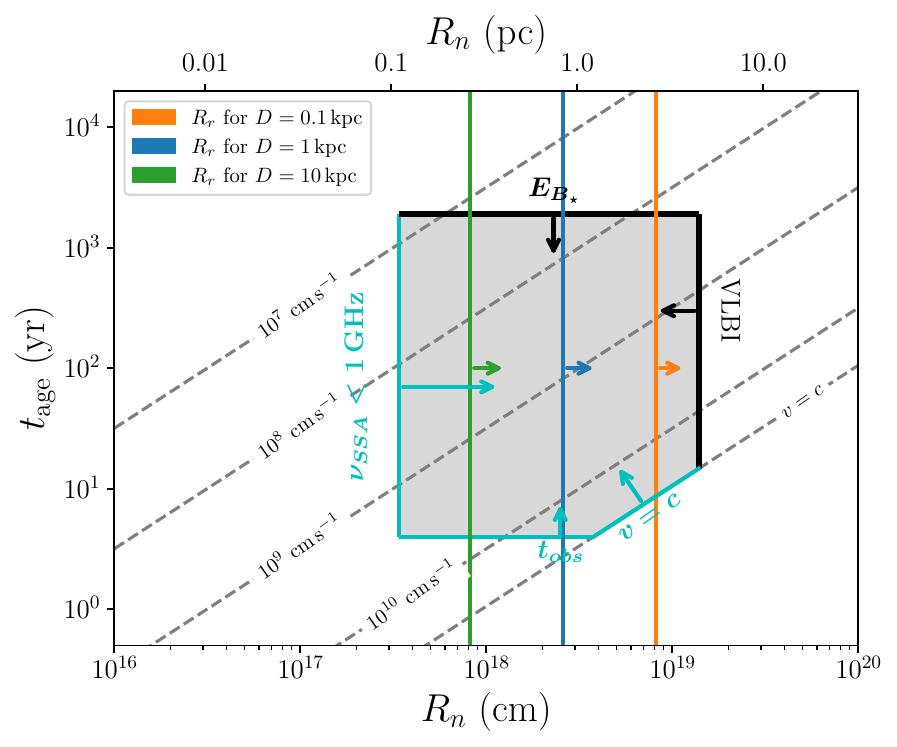}
    \caption{Constraint on the radius of the PRS source using the projected size of the screen $R_{r}$. The figure is adapted from \cite{2023ApJ...958L..19B}, after correcting for the radius by dividing all limits by 2.}
    \label{fig:param_phase_space_plot}
\end{figure}

\textbf{ We obtain the observed modulation index $\bm{m_{\rm{obs}}=0.23}$. The observed modulation of the PRS flux is less than that expected from this region of the sky, i.e. $\bm{m_{\rm{obs}}<m_{\rm{exp}}}$.} This observation can be interpreted in the following ways:

\begin{itemize}
    \item \textbf{All the variability observed is due to scintillation, and the pyNE2001 estimate of the transition frequency is accurate}. However, the observed flux modulation is lesser than the expected modulation. This change in modulation can be explained if the size of the source $\theta_{s}$ is slightly larger than the size of the screen $\theta_{r}$, reducing the modulation index by a factor of $(\theta_{r}/\theta_{s})^{7/6}$, and increasing refractive timescale by a factor of $\theta_{s}/\theta_{r}$ \citep[see][for details]{1998MNRAS.294..307W}. Table \ref{tab:scintillation} summarizes the possible constraints on the size of the source ($R_{s}=\theta_{s}d_{\rm{lum}}/2$), and the modified timescale of the scintillation, for screens at a distance of 0.1\,kpc, 1\,kpc and 10\,kpc.  The projected screen radius ($R_{r}$) values listed in Table \ref{tab:scintillation} can be quoted as a conservative lower limit of the size of the source. These are shown in Figure \ref{fig:param_phase_space_plot} for different distances of the screen from the observer in the $t_{\rm{age}}$--$R_{n}$ phase space plot, adapted from \cite{2023ApJ...958L..19B}. 
    
    \item \textbf{All the variability observed is due to scintillation, but the pyNE2001 model estimate of the transition frequency is not accurate.} If this is the case, we can estimate the transition frequency required to produce the observed modulation. Table \ref{tab:scintillation} lists the expected transition frequency, $\nu_{0}$ which can produce the observed modulation index at $\nu=1.4\,$GHz. 

    \item \textbf{All the variability observed is intrinsic to the source}. This implies that the size of the source $\theta_{s}$ is larger than the size of the screen $\theta_{r}$ and therefore, refractive scintillation (which assumes a point source) is not relevant here. The projected size of the screen $R_{r}$ tabulated in Table \ref{tab:scintillation} can be used as a conservative lower limit of the size of the source in this case.
\end{itemize}

At this point, we also note that there is a possibility of both intrinsic variability and scintillation to contribute to the observed variability. It is not trivial to separate the two without independent constraints on the size of the source and the properties of the scattering screen. 

For completeness, we also repeat this analysis after removing the expected decay from the magnetar wind nebula model (Section~\ref{subsec:PWN}). The inferred size and timescale constraints remain broadly consistent in both cases.

\begin{deluxetable*}{|c|c|c|c|c|c|c|c|c|c|c|c|c|c|c|}
\label{tab:scintillation}
\tablecaption{Scintillation analysis and predictions at $\approx$1.4\,GHz (the observation frequency used is a mean of all the detections from Table \ref{tab:comprehensive_tab} i.e. $\nu\approx1.4$\,GHz)}
\tablehead{
\multirow{4}{*}{Dataset} & \multirow{4}{*}{\shortstack{$t_{r}$\\(days)}} & \multirow{4}{*}{$m_{\rm{exp}}$} & \multirow{4}{*}{$m_{\rm{obs}}$} &  \multicolumn{9}{c|}{Characteristic sizes} & \multirow{4}{*}{\shortstack{modified\\$t_{r}$\\ [1em] (days)}} & \multirow{4}{*}{\shortstack{modified\\$\nu_{0}$\\[1em] (\,GHz)}}\\
\cline{5-13}
 & & & & \multicolumn{3}{c|}{$D=0.1$\,kpc} & \multicolumn{3}{c|}{$D=1$\,kpc} & \multicolumn{3}{c|}{$D=10$\,kpc} & &\\
\cline{5-13}
 & & & & & & & & & & & & & & \\
 & & & & \shortstack{$\theta_{F}$\\($\mu$as)} & \shortstack{$R_{r}$\\(pc)} & \shortstack{$R_{s}$\\(pc)} & \shortstack{$\theta_{F}$\\($\mu$as)} & \shortstack{$R_{r}$\\(pc)} & \shortstack{$R_{s}$\\(pc)} & \shortstack{$\theta_{F}$\\($\mu$as)} & \shortstack{$R_{r}$\\(pc)} & \shortstack{$R_{s}$\\(pc)} & &}
\startdata
 Data as it is & \multirow{2}{*}{10.47} & \multirow{2}{*}{0.29} & 0.23 & \multirow{2}{*}{21.44
} & \multirow{2}{*}{5.30} & 6.48 & \multirow{2}{*}{6.78} & \multirow{2}{*}{1.68} & 2.05 & \multirow{2}{*}{2.14} & \multirow{2}{*}{0.53} & 0.65 & 12.80 & 18.93 \\
 \cline{1-1}
 \cline{4-4}
 \cline{7-7}
 \cline{10-10}
 \cline{13-15}
  Removed best fit decay & & & 0.22 & & & 6.80 & & & 2.15 & & & 0.68 & 13.43 & 20.92 \\
\enddata
\end{deluxetable*}

\subsection{Hypernebula Model}
Another model proposed to explain the PRSs associated with FRBs is the hypernebula model \citep{2022ApJ...937....5S}. In this model, mass transfer from a companion star onto a massive black hole via Roche lobe overflow drives powerful disk winds. These winds inflate a bubble of relativistic electrons that emit synchrotron radiation. Two key timescales characterize the model: the timescale over which the wind expands freely into the CSM, $t_{\rm{free}}$; and $t_{\rm{active}}\sim10$--$10^{6}$\,years, the total duration of the accretion phase \citep{2022ApJ...937....5S}. The relatively modest evolution in burst DM over time \citep{2022Natur.606..873N,2023Sci...380..599A} suggests that the system is currently at $t < t_{\rm{free}}$. However, \cite{2021ApJ...923L..17Z} show a decreasing trend in the burst DM. \citep{2022ApJ...937....5S} applied the hypernebula model specifically to \ourfrb, using a set of best-fit model parameters (e.g., age, mass accretion rate, ambient density) chosen to match the source’s known properties at the time. For those specific parameters, their model predicts a gradual increase in radio flux of the PRS at GHz frequencies over decadal timescales—this is explicitly stated in Section 4.2 of their paper and shown in their model curves \citep[see Figures 5 and 6 in][]{2022ApJ...937....5S}. Whereas, we see a slow decay in the flux over the span of $\sim4$\,years. This is in contrast to our observations, indicating that the best-fit hypernebula model using data from the first 2 years since detection is inconsistent with the recent observed behavior of the PRS associated with \ourfrb. 

\subsection{AGN or IMBH Origin for the Persistent Radio Source}
\label{subsec:agn_origin}

An alternative scenario is that the persistent radio source (PRS) associated with \ourfrb\ is physically unrelated to the FRB engine, and instead arises from accretion onto a massive black hole as discussed by \citet{2023Sci...380..599A}. In this picture, the PRS represents compact synchrotron emission from a radio-loud, low-Eddington active galactic nucleus (AGN), while the FRB originates from a separate compact object—such as a magnetar—embedded within the same nuclear environment.

Recent work by \citet{2024ApJ...973..133D} has systematically compared the radio spectra of persistent sources in nearby dwarf galaxies—including known low-mass AGNs—with those of FRB-associated PRSs, including \ourfrb\ and FRB~20121102A. They find that several AGNs in dwarf hosts exhibit flat or slightly inverted radio spectra and luminosities consistent with both FRB PRSs, suggesting that an AGN origin cannot be ruled out based on spectral properties alone.

For \ourfrb, the persistent source exhibits a flat spectrum, compact morphology (unresolved on $\sim$10~pc scales), and high radio luminosity ($L_\nu \sim 3 \times 10^{29}\,\mathrm{erg\,s^{-1}\,Hz^{-1}}$), all of which are consistent with a jet-dominated AGN powered by an intermediate mass black hole (IMBH). Importantly, recent \textit{XMM-Newton} and \textit{Chandra} observations place an upper limit on the X-ray luminosity of $L_X \lesssim 9 \times 10^{42}$\,erg\,s$^{-1}$ (3$\sigma$) \citep{2025ApJ...983..116Y}, which yields a radio-to-X-ray luminosity ratio of $\log(R_X) \gtrsim -3$. This value lies within the regime of radio-loud AGNs \citep{2003ApJ...583..145T}, and is broadly consistent with radiatively inefficient accretion flows.

A similar interpretation has been proposed for the PRS associated with FRB~20121102A. \citet{2025PASP..137h4202B} argued that its long-term flux stability, compact size ($<0.35$\,pc), and placement on the radio-loud fundamental plane of black hole activity can be explained by a low-Eddington AGN powered by an IMBH. Given the morphological and spectral similarities between the two sources, a comparable AGN origin for the PRS of \ourfrb\ remains observationally viable. However, as with FRB~20121102A, this model requires that the FRB bursts themselves arise from a distinct, co-located source within the broader accretion environment.

Further high-resolution radio monitoring and deep X-ray observations will be essential to distinguish between AGN and magnetar wind nebula interpretations and to assess whether FRB sources and persistent emitters are causally connected or merely cohabit similar extreme environments.

\section{Conclusion}
\label{sec:conclusion}
In this paper, we present broadband radio observations ($0.3$–-$10$\,GHz) of the persistent radio source (PRS) associated with \ourfrb. We analyze the data under different proposed models and comment on the physics responsible for the observed emission. The findings are summarized below.
\begin{itemize}
    \item We observe a slow decay in the brightness of the PRS with time across all frequency bands. There are episodes of brightening and dimming seen in the 1.5\,GHz and 3\,GHz data.
    \item The first low frequency ($<1\,$GHz) observations of this PRS hint at a spectral break between $\approx0.7$\,GHz and $\approx1.3$\,GHz, which can be attributed to synchrotron self-absorption. It is not possible to estimate the self-absorption frequency using only the upper limits from the low-frequency observations presented here.
    \item We fit the decaying light curves at all frequencies using the decay equation described in Equation \ref{eq:joint_fit}, derived from the magnetar wind nebula model. This resulted in an estimate of the powerlaw index for energy injection $\alpha=1.77^{+0.07}_{-0.08}$ and the age of the magnetar is $t_{\rm{age}}$, is $52^{+16}_{-10}$\,years. These values are in agreement with constraints from previous observations of the PRS \citep{2023ApJ...958L..19B,2021ApJ...923L..17Z, 2024arXiv241219358B}.
    \item We also comment on the contribution of scintillation to the variability of the PRS at 1.5\,GHz. Through this analysis,  a conservative lower limit on the size of the source $>0.53$\,pc (for a screen at 10\,kpc) is derived (assuming that NE2001 accurately models the transition frequency and that the observed variability is due to scintillation). This limit is consistent with constraints from VLBI observations and low-frequency non-detection. We also note that the variability could be a combination of scintillation and intrinsic variability, and it is not trivial to distinguish between the two.
    \item Finally, we discuss the observations under the purview of the best-fit hypernebula model for the PRS associated with \ourfrb\ presented in \citep{2022ApJ...937....5S}. This best-fit model predicts a rise in flux with time at 1.5\,GHz, which does not agree with the observations presented here.
\end{itemize}

Ongoing monitoring of known PRSs, along with future detections of new PRSs associated with FRBs, will be crucial in further constraining the emission mechanisms and understanding the role of the local environment in the persistent emission.

\software{CASA \citep{2022PASP..134k4501C}, CASA CAPTURE \citep{2021ExA....51...95K}, emcee \citep{2013PASP..125..306F}, ChainConsumer \citep{Hinton2016}, Astropy \citep{astropy:2013, astropy:2018, astropy:2022}, NumPy \citep{harris2020array}, SciPy \citep{2020SciPy-NMeth}, Matplotlib \citep{Hunter:2007}}

\begin{acknowledgements}
\renewcommand\linenumberfont{\color{white}}

We thank the staff of the GMRT that made these observations possible. GMRT is run by the National Centre for Radio Astrophysics of the Tata Institute of Fundamental Research. The National Radio Astronomy Observatory is a facility of the National Science Foundation operated under cooperative agreement by Associated Universities, Inc. M.B is a McWilliams fellow and an International Astronomical Union Gruber fellow. M.B. also receives support from the McWilliams seed grant. Finally, we gratefully acknowledge the insightful discussions held during the FRB Frontiers Meeting (March 3–5, 2025), which contributed significantly to the development of this work \citep{2025NatAs.tmp..149B}. A.B thanks Deeepak Eappachen for useful discussions.
\end{acknowledgements}


\bibliography{references}{}

@ARTICLE{2022A&ARv..30....2P,
       author = {{Petroff}, E. and {Hessels}, J.~W.~T. and {Lorimer}, D.~R.},
        title = "{Fast radio bursts at the dawn of the 2020s}",
      journal = {\aapr},
         year = 2022,
        month = dec,
       volume = {30},
       number = {1},
          eid = {2},
        pages = {2},
          doi = {10.1007/s00159-022-00139-w},
archivePrefix = {arXiv},
       eprint = {2107.10113},
 primaryClass = {astro-ph.HE},
       adsurl = {https://ui.adsabs.harvard.edu/abs/2022A&ARv..30....2P}
}

@ARTICLE{2025NatAs.tmp..149B,
       author = {{Bhardwaj}, Mohit and {McLaughlin}, Maura},
        title = "{Peering beyond radio flashes with multi-messenger insight into FRB engines}",
      journal = {Nature Astronomy},
         year = 2025,
        month = jul,
          doi = {10.1038/s41550-025-02611-w},
       adsurl = {https://ui.adsabs.harvard.edu/abs/2025NatAs.tmp..149B}
}

@ARTICLE{2023ApJ...959...89Z,
       author = {{Zhang}, Xian and {Yu}, Wenfei and {Law}, Casey and {Li}, Di and {Chatterjee}, Shami and {Demorest}, Paul and {Yan}, Zhen and {Niu}, Chenhui and {Aggarwal}, Kshitij and {Anna-Thomas}, Reshma and {Burke-Spolaor}, Sarah and {Connor}, Liam and {Tsai}, Chao-Wei and {Zhu}, Weiwei and {Luo}, Gan},
        title = "{Temporal and Spectral Properties of the Persistent Radio Source Associated with FRB 20190520B with the VLA}",
      journal = {\apj},
         year = 2023,
        month = dec,
       volume = {959},
       number = {2},
          eid = {89},
        pages = {89},
          doi = {10.3847/1538-4357/ad0545},
archivePrefix = {arXiv},
       eprint = {2307.16355},
 primaryClass = {astro-ph.HE},
       adsurl = {https://ui.adsabs.harvard.edu/abs/2023ApJ...959...89Z}
}

@ARTICLE{2023ApJ...958L..19B,
       author = {{Bhandari}, Shivani and {Marcote}, Benito and {Sridhar}, Navin and {Eftekhari}, Tarraneh and {Hessels}, Jason W.~T. and {Hewitt}, Dant{\'e} M. and {Kirsten}, Franz and {Ould-Boukattine}, Omar S. and {Paragi}, Zsolt and {Snelders}, Mark P.},
        title = "{Constraints on the Persistent Radio Source Associated with FRB 20190520B Using the European VLBI Network}",
      journal = {\apjl},
         year = 2023,
        month = dec,
       volume = {958},
       number = {2},
          eid = {L19},
        pages = {L19},
          doi = {10.3847/2041-8213/ad083f},
archivePrefix = {arXiv},
       eprint = {2308.12801},
 primaryClass = {astro-ph.HE},
       adsurl = {https://ui.adsabs.harvard.edu/abs/2023ApJ...958L..19B}
}

@ARTICLE{2023Sci...380..599A,
       author = {{Anna-Thomas}, Reshma and {Connor}, Liam and {Dai}, Shi and {Feng}, Yi and {Burke-Spolaor}, Sarah and {Beniamini}, Paz and {Yang}, Yuan-Pei and {Zhang}, Yong-Kun and {Aggarwal}, Kshitij and {Law}, Casey J. and {Li}, Di and {Niu}, Chenhui and {Chatterjee}, Shami and {Cruces}, Marilyn and {Duan}, Ran and {Filipovic}, Miroslav D. and {Hobbs}, George and {Lynch}, Ryan S. and {Miao}, Chenchen and {Niu}, Jiarui and {Ocker}, Stella K. and {Tsai}, Chao-Wei and {Wang}, Pei and {Xue}, Mengyao and {Yao}, Ju-Mei and {Yu}, Wenfei and {Zhang}, Bing and {Zhang}, Lei and {Zhu}, Shiqiang and {Zhu}, Weiwei},
        title = "{Magnetic field reversal in the turbulent environment around a repeating fast radio burst}",
      journal = {Science},
         year = 2023,
        month = may,
       volume = {380},
       number = {6645},
        pages = {599-603},
          doi = {10.1126/science.abo6526},
archivePrefix = {arXiv},
       eprint = {2202.11112},
 primaryClass = {astro-ph.HE},
       adsurl = {https://ui.adsabs.harvard.edu/abs/2023Sci...380..599A}
}

@ARTICLE{2022Natur.606..873N,
       author = {{Niu}, C. -H. and {Aggarwal}, K. and {Li}, D. and {Zhang}, X. and {Chatterjee}, S. and {Tsai}, C. -W. and {Yu}, W. and {Law}, C.~J. and {Burke-Spolaor}, S. and {Cordes}, J.~M. and {Zhang}, Y. -K. and {Ocker}, S.~K. and {Yao}, J. -M. and {Wang}, P. and {Feng}, Y. and {Niino}, Y. and {Bochenek}, C. and {Cruces}, M. and {Connor}, L. and {Jiang}, J. -A. and {Dai}, S. and {Luo}, R. and {Li}, G. -D. and {Miao}, C. -C. and {Niu}, J. -R. and {Anna-Thomas}, R. and {Sydnor}, J. and {Stern}, D. and {Wang}, W. -Y. and {Yuan}, M. and {Yue}, Y. -L. and {Zhou}, D. -J. and {Yan}, Z. and {Zhu}, W. -W. and {Zhang}, B.},
        title = "{A repeating fast radio burst associated with a persistent radio source}",
      journal = {\nat},
         year = 2022,
        month = jun,
       volume = {606},
       number = {7916},
        pages = {873-877},
          doi = {10.1038/s41586-022-04755-5},
archivePrefix = {arXiv},
       eprint = {2110.07418},
 primaryClass = {astro-ph.HE},
       adsurl = {https://ui.adsabs.harvard.edu/abs/2022Natur.606..873N}
}

@ARTICLE{2019PhR...821....1P,
       author = {{Platts}, E. and {Weltman}, A. and {Walters}, A. and {Tendulkar}, S.~P. and {Gordin}, J.~E.~B. and {Kandhai}, S.},
        title = "{A living theory catalogue for fast radio bursts}",
      journal = {\physrep},
         year = 2019,
        month = aug,
       volume = {821},
        pages = {1-27},
          doi = {10.1016/j.physrep.2019.06.003},
archivePrefix = {arXiv},
       eprint = {1810.05836},
 primaryClass = {astro-ph.HE},
       adsurl = {https://ui.adsabs.harvard.edu/abs/2019PhR...821....1P}
}

@ARTICLE{2025ApJ...983..116Y,
       author = {{Yan}, Zhen and {Yu}, Wenfei and {Page}, Kim L. and {Lin}, Jie and {Li}, Di and {Niu}, Chenhui and {Law}, Casey and {Zhang}, Bing and {Chatterjee}, Shami and {Zhang}, Xian and {Anna-Thomas}, Reshma},
        title = "{Simultaneous Multiwavelength Observations of the Repeating Fast Radio Burst FRB 20190520B with Swift and FAST}",
      journal = {\apj},
         year = 2025,
        month = apr,
       volume = {983},
       number = {2},
          eid = {116},
        pages = {116},
          doi = {10.3847/1538-4357/adbef0},
archivePrefix = {arXiv},
       eprint = {2402.12084},
 primaryClass = {astro-ph.HE},
       adsurl = {https://ui.adsabs.harvard.edu/abs/2025ApJ...983..116Y}
}

@ARTICLE{2003ApJ...583..145T,
       author = {{Terashima}, Yuichi and {Wilson}, Andrew S.},
        title = "{Chandra Snapshot Observations of Low-Luminosity Active Galactic Nuclei with a Compact Radio Source}",
      journal = {\apj},
         year = 2003,
        month = jan,
       volume = {583},
       number = {1},
        pages = {145-158},
          doi = {10.1086/345339},
archivePrefix = {arXiv},
       eprint = {astro-ph/0209607},
 primaryClass = {astro-ph},
       adsurl = {https://ui.adsabs.harvard.edu/abs/2003ApJ...583..145T}
}

@ARTICLE{2020Natur.587...59B,
       author = {{Bochenek}, C.~D. and {Ravi}, V. and {Belov}, K.~V. and {Hallinan}, G. and {Kocz}, J. and {Kulkarni}, S.~R. and {McKenna}, D.~L.},
        title = "{A fast radio burst associated with a Galactic magnetar}",
      journal = {\nat},
         year = 2020,
        month = nov,
       volume = {587},
       number = {7832},
        pages = {59-62},
          doi = {10.1038/s41586-020-2872-x},
archivePrefix = {arXiv},
       eprint = {2005.10828},
 primaryClass = {astro-ph.HE},
       adsurl = {https://ui.adsabs.harvard.edu/abs/2020Natur.587...59B}
}

@ARTICLE{2020Natur.587...54C,
       author = {{CHIME/FRB Collaboration} and {Andersen}, B.~C. and {Bandura}, K.~M. and {Bhardwaj}, M. and {Bij}, A. and {Boyce}, M.~M. and {Boyle}, P.~J. and {Brar}, C. and {Cassanelli}, T. and {Chawla}, P. and {Chen}, T. and {Cliche}, J. -F. and {Cook}, A. and {Cubranic}, D. and {Curtin}, A.~P. and {Denman}, N.~T. and {Dobbs}, M. and {Dong}, F.~Q. and {Fandino}, M. and {Fonseca}, E. and {Gaensler}, B.~M. and {Giri}, U. and {Good}, D.~C. and {Halpern}, M. and {Hill}, A.~S. and {Hinshaw}, G.~F. and {H{\"o}fer}, C. and {Josephy}, A. and {Kania}, J.~W. and {Kaspi}, V.~M. and {Landecker}, T.~L. and {Leung}, C. and {Li}, D.~Z. and {Lin}, H. -H. and {Masui}, K.~W. and {McKinven}, R. and {Mena-Parra}, J. and {Merryfield}, M. and {Meyers}, B.~W. and {Michilli}, D. and {Milutinovic}, N. and {Mirhosseini}, A. and {M{\"u}nchmeyer}, M. and {Naidu}, A. and {Newburgh}, L.~B. and {Ng}, C. and {Patel}, C. and {Pen}, U. -L. and {Pinsonneault-Marotte}, T. and {Pleunis}, Z. and {Quine}, B.~M. and {Rafiei-Ravandi}, M. and {Rahman}, M. and {Ransom}, S.~M. and {Renard}, A. and {Sanghavi}, P. and {Scholz}, P. and {Shaw}, J.~R. and {Shin}, K. and {Siegel}, S.~R. and {Singh}, S. and {Smegal}, R.~J. and {Smith}, K.~M. and {Stairs}, I.~H. and {Tan}, C.~M. and {Tendulkar}, S.~P. and {Tretyakov}, I. and {Vanderlinde}, K. and {Wang}, H. and {Wulf}, D. and {Zwaniga}, A.~V.},
        title = "{A bright millisecond-duration radio burst from a Galactic magnetar}",
      journal = {\nat},
         year = 2020,
        month = nov,
       volume = {587},
       number = {7832},
        pages = {54-58},
          doi = {10.1038/s41586-020-2863-y},
archivePrefix = {arXiv},
       eprint = {2005.10324},
 primaryClass = {astro-ph.HE},
       adsurl = {https://ui.adsabs.harvard.edu/abs/2020Natur.587...54C}
}

@ARTICLE{2017Natur.541...58C,
       author = {{Chatterjee}, S. and {Law}, C.~J. and {Wharton}, R.~S. and {Burke-Spolaor}, S. and {Hessels}, J.~W.~T. and {Bower}, G.~C. and {Cordes}, J.~M. and {Tendulkar}, S.~P. and {Bassa}, C.~G. and {Demorest}, P. and {Butler}, B.~J. and {Seymour}, A. and {Scholz}, P. and {Abruzzo}, M.~W. and {Bogdanov}, S. and {Kaspi}, V.~M. and {Keimpema}, A. and {Lazio}, T.~J.~W. and {Marcote}, B. and {McLaughlin}, M.~A. and {Paragi}, Z. and {Ransom}, S.~M. and {Rupen}, M. and {Spitler}, L.~G. and {van Langevelde}, H.~J.},
        title = "{A direct localization of a fast radio burst and its host}",
      journal = {\nat},
         year = 2017,
        month = jan,
       volume = {541},
       number = {7635},
        pages = {58-61},
          doi = {10.1038/nature20797},
archivePrefix = {arXiv},
       eprint = {1701.01098},
 primaryClass = {astro-ph.HE},
       adsurl = {https://ui.adsabs.harvard.edu/abs/2017Natur.541...58C}
}

@ARTICLE{2022PASP..134k4501C,
       author = {{CASA Team} and {Bean}, Ben and {Bhatnagar}, Sanjay and {Castro}, Sandra and {Donovan Meyer}, Jennifer and {Emonts}, Bjorn and {Garcia}, Enrique and {Garwood}, Robert and {Golap}, Kumar and {Gonzalez Villalba}, Justo and {Harris}, Pamela and {Hayashi}, Yohei and {Hoskins}, Josh and {Hsieh}, Mingyu and {Jagannathan}, Preshanth and {Kawasaki}, Wataru and {Keimpema}, Aard and {Kettenis}, Mark and {Lopez}, Jorge and {Marvil}, Joshua and {Masters}, Joseph and {McNichols}, Andrew and {Mehringer}, David and {Miel}, Renaud and {Moellenbrock}, George and {Montesino}, Federico and {Nakazato}, Takeshi and {Ott}, Juergen and {Petry}, Dirk and {Pokorny}, Martin and {Raba}, Ryan and {Rau}, Urvashi and {Schiebel}, Darrell and {Schweighart}, Neal and {Sekhar}, Srikrishna and {Shimada}, Kazuhiko and {Small}, Des and {Steeb}, Jan-Willem and {Sugimoto}, Kanako and {Suoranta}, Ville and {Tsutsumi}, Takahiro and {van Bemmel}, Ilse M. and {Verkouter}, Marjolein and {Wells}, Akeem and {Xiong}, Wei and {Szomoru}, Arpad and {Griffith}, Morgan and {Glendenning}, Brian and {Kern}, Jeff},
        title = "{CASA, the Common Astronomy Software Applications for Radio Astronomy}",
      journal = {\pasp},
         year = 2022,
        month = nov,
       volume = {134},
       number = {1041},
          eid = {114501},
        pages = {114501},
          doi = {10.1088/1538-3873/ac9642},
archivePrefix = {arXiv},
       eprint = {2210.02276},
 primaryClass = {astro-ph.IM},
       adsurl = {https://ui.adsabs.harvard.edu/abs/2022PASP..134k4501C}
}

@ARTICLE{2021ExA....51...95K,
       author = {{Kale}, Ruta and {Ishwara-Chandra}, C.~H.},
        title = "{CAPTURE: a continuum imaging pipeline for the uGMRT}",
      journal = {Experimental Astronomy},
         year = 2021,
        month = feb,
       volume = {51},
       number = {1},
        pages = {95-108},
          doi = {10.1007/s10686-020-09677-6},
archivePrefix = {arXiv},
       eprint = {2010.00196},
 primaryClass = {astro-ph.IM},
       adsurl = {https://ui.adsabs.harvard.edu/abs/2021ExA....51...95K}
}

@ARTICLE{1998MNRAS.294..307W,
       author = {{Walker}, Mark A.},
        title = "{Interstellar scintillation of compact extragalactic radio sources}",
      journal = {\mnras},
         year = 1998,
        month = feb,
       volume = {294},
        pages = {307-311},
          doi = {10.1111/j.1365-8711.1998.01238.x},
       adsurl = {https://ui.adsabs.harvard.edu/abs/1998MNRAS.294..307W}
}

@ARTICLE{1992RSPTA.341..151N,
       author = {{Narayan}, Ramesh},
        title = "{The Physics of Pulsar Scintillation}",
      journal = {Philosophical Transactions of the Royal Society of London Series A},
         year = 1992,
        month = oct,
       volume = {341},
       number = {1660},
        pages = {151-165},
          doi = {10.1098/rsta.1992.0090},
       adsurl = {https://ui.adsabs.harvard.edu/abs/1992RSPTA.341..151N}
}

@ARTICLE{2002astro.ph..7156C,
       author = {{Cordes}, J.~M. and {Lazio}, T.~J.~W.},
        title = "{NE2001.I. A New Model for the Galactic Distribution of Free Electrons and its Fluctuations}",
      journal = {arXiv e-prints},
         year = 2002,
        month = jul,
          eid = {astro-ph/0207156},
        pages = {astro-ph/0207156},
          doi = {10.48550/arXiv.astro-ph/0207156},
archivePrefix = {arXiv},
       eprint = {astro-ph/0207156},
 primaryClass = {astro-ph},
       adsurl = {https://ui.adsabs.harvard.edu/abs/2002astro.ph..7156C}
}

@ARTICLE{2018ApJ...868L...4M,
       author = {{Margalit}, Ben and {Metzger}, Brian D.},
        title = "{A Concordance Picture of FRB 121102 as a Flaring Magnetar Embedded in a Magnetized Ion-Electron Wind Nebula}",
      journal = {\apjl},
         year = 2018,
        month = nov,
       volume = {868},
       number = {1},
          eid = {L4},
        pages = {L4},
          doi = {10.3847/2041-8213/aaedad},
archivePrefix = {arXiv},
       eprint = {1808.09969},
 primaryClass = {astro-ph.HE},
       adsurl = {https://ui.adsabs.harvard.edu/abs/2018ApJ...868L...4M}
}

@ARTICLE{2013PASP..125..306F,
       author = {{Foreman-Mackey}, Daniel and {Hogg}, David W. and {Lang}, Dustin and {Goodman}, Jonathan},
        title = "{emcee: The MCMC Hammer}",
      journal = {\pasp},
         year = 2013,
        month = mar,
       volume = {125},
       number = {925},
        pages = {306},
          doi = {10.1086/670067},
archivePrefix = {arXiv},
       eprint = {1202.3665},
 primaryClass = {astro-ph.IM},
       adsurl = {https://ui.adsabs.harvard.edu/abs/2013PASP..125..306F}
}

@ARTICLE{Hinton2016,
      author = {{Hinton}, S.~R.},
       title = "{ChainConsumer}",
     journal = {The Journal of Open Source Software},
        year = 2016,
       month = aug,
      volume = 1,
         eid = {00045},
       pages = {00045},
         doi = {10.21105/joss.00045},
      adsurl = {http://adsabs.harvard.edu/abs/2016JOSS....1...45H},
}

@ARTICLE{2022ApJ...937....5S,
       author = {{Sridhar}, Navin and {Metzger}, Brian D.},
        title = "{Radio Nebulae from Hyperaccreting X-Ray Binaries as Common-envelope Precursors and Persistent Counterparts of Fast Radio Bursts}",
      journal = {\apj},
         year = 2022,
        month = sep,
       volume = {937},
       number = {1},
          eid = {5},
        pages = {5},
          doi = {10.3847/1538-4357/ac8a4a},
archivePrefix = {arXiv},
       eprint = {2206.10486},
 primaryClass = {astro-ph.HE},
       adsurl = {https://ui.adsabs.harvard.edu/abs/2022ApJ...937....5S}
}

@ARTICLE{2024ApJ...969..145C,
       author = {{CHIME/FRB Collaboration} and {Amiri}, Mandana and {Andersen}, Bridget C. and {Andrew}, Shion and {Bandura}, Kevin and {Bhardwaj}, Mohit and {Boyle}, P.~J. and {Brar}, Charanjot and {Breitman}, Daniela and {Cassanelli}, Tomas and {Chawla}, Pragya and {Cook}, Amanda M. and {Curtin}, Alice P. and {Dobbs}, Matt and {Dong}, Fengqiu Adam and {Eadie}, Gwendolyn and {Fonseca}, Emmanuel and {Gaensler}, B.~M. and {Giri}, Utkarsh and {Herrera-Martin}, Antonio and {Hopkins}, Hans and {Ibik}, Adaeze L. and {Joseph}, Ronniy C. and {Kaczmarek}, J.~F. and {Kader}, Zarif and {Kaspi}, Victoria M. and {Lanman}, Adam E. and {Lazda}, Mattias and {Leung}, Calvin and {Liu}, Siqi and {Masui}, Kiyoshi W. and {McKinven}, Ryan and {Mena-Parra}, Juan and {Merryfield}, Marcus and {Michilli}, Daniele and {Ng}, Cherry and {Nimmo}, Kenzie and {Noble}, Gavin and {Pandhi}, Ayush and {Patel}, Chitrang and {Pearlman}, Aaron B. and {Pen}, Ue-Li and {Petroff}, Emily and {Pleunis}, Ziggy and {Rafiei-Ravandi}, Masoud and {Rahman}, Mubdi and {Ransom}, Scott M. and {Sand}, Ketan R. and {Scholz}, Paul and {Shah}, Vishwangi and {Shin}, Kaitlyn and {Shpunarska}, Yuliya and {Siegel}, Seth R. and {Smith}, Kendrick and {Stairs}, Ingrid and {Stenning}, David C. and {Vanderlinde}, Keith and {Wang}, Haochen and {White}, Henry and {Wulf}, Dallas},
        title = "{Updating the First CHIME/FRB Catalog of Fast Radio Bursts with Baseband Data}",
      journal = {\apj},
         year = 2024,
        month = jul,
       volume = {969},
       number = {2},
          eid = {145},
        pages = {145},
          doi = {10.3847/1538-4357/ad464b},
archivePrefix = {arXiv},
       eprint = {2311.00111},
 primaryClass = {astro-ph.HE},
       adsurl = {https://ui.adsabs.harvard.edu/abs/2024ApJ...969..145C}
}

@ARTICLE{2021ApJ...908L..12T,
       author = {{Tendulkar}, Shriharsh P. and {Gil de Paz}, Armando and {Kirichenko}, Aida Yu. and {Hessels}, Jason W.~T. and {Bhardwaj}, Mohit and {{\'A}vila}, Fernando and {Bassa}, Cees and {Chawla}, Pragya and {Fonseca}, Emmanuel and {Kaspi}, Victoria M. and {Keimpema}, Aard and {Kirsten}, Franz and {Lazio}, T. Joseph W. and {Marcote}, Benito and {Masui}, Kiyoshi and {Nimmo}, Kenzie and {Paragi}, Zsolt and {Rahman}, Mubdi and {Pay{\'a}}, Daniel Reverte and {Scholz}, Paul and {Stairs}, Ingrid},
        title = "{The 60 pc Environment of FRB 20180916B}",
      journal = {\apjl},
         year = 2021,
        month = feb,
       volume = {908},
       number = {1},
          eid = {L12},
        pages = {L12},
          doi = {10.3847/2041-8213/abdb38},
archivePrefix = {arXiv},
       eprint = {2011.03257},
 primaryClass = {astro-ph.HE},
       adsurl = {https://ui.adsabs.harvard.edu/abs/2021ApJ...908L..12T}
}

@ARTICLE{2020Natur.577..190M,
       author = {{Marcote}, B. and {Nimmo}, K. and {Hessels}, J.~W.~T. and {Tendulkar}, S.~P. and {Bassa}, C.~G. and {Paragi}, Z. and {Keimpema}, A. and {Bhardwaj}, M. and {Karuppusamy}, R. and {Kaspi}, V.~M. and {Law}, C.~J. and {Michilli}, D. and {Aggarwal}, K. and {Andersen}, B. and {Archibald}, A.~M. and {Bandura}, K. and {Bower}, G.~C. and {Boyle}, P.~J. and {Brar}, C. and {Burke-Spolaor}, S. and {Butler}, B.~J. and {Cassanelli}, T. and {Chawla}, P. and {Demorest}, P. and {Dobbs}, M. and {Fonseca}, E. and {Giri}, U. and {Good}, D.~C. and {Gourdji}, K. and {Josephy}, A. and {Kirichenko}, A. Yu. and {Kirsten}, F. and {Landecker}, T.~L. and {Lang}, D. and {Lazio}, T.~J.~W. and {Li}, D.~Z. and {Lin}, H. -H. and {Linford}, J.~D. and {Masui}, K. and {Mena-Parra}, J. and {Naidu}, A. and {Ng}, C. and {Patel}, C. and {Pen}, U. -L. and {Pleunis}, Z. and {Rafiei-Ravandi}, M. and {Rahman}, M. and {Renard}, A. and {Scholz}, P. and {Siegel}, S.~R. and {Smith}, K.~M. and {Stairs}, I.~H. and {Vanderlinde}, K. and {Zwaniga}, A.~V.},
        title = "{A repeating fast radio burst source localized to a nearby spiral galaxy}",
      journal = {\nat},
         year = 2020,
        month = jan,
       volume = {577},
       number = {7789},
        pages = {190-194},
          doi = {10.1038/s41586-019-1866-z},
archivePrefix = {arXiv},
       eprint = {2001.02222},
 primaryClass = {astro-ph.HE},
       adsurl = {https://ui.adsabs.harvard.edu/abs/2020Natur.577..190M}
}

@ARTICLE{2007Sci...318..777L,
       author = {{Lorimer}, D.~R. and {Bailes}, M. and {McLaughlin}, M.~A. and {Narkevic}, D.~J. and {Crawford}, F.},
        title = "{A Bright Millisecond Radio Burst of Extragalactic Origin}",
      journal = {Science},
         year = 2007,
        month = nov,
       volume = {318},
       number = {5851},
        pages = {777},
          doi = {10.1126/science.1147532},
archivePrefix = {arXiv},
       eprint = {0709.4301},
 primaryClass = {astro-ph},
       adsurl = {https://ui.adsabs.harvard.edu/abs/2007Sci...318..777L}
}

@ARTICLE{2021ApJS..257...59C,
       author = {{CHIME/FRB Collaboration} and {Amiri}, Mandana and {Andersen}, Bridget C. and {Bandura}, Kevin and {Berger}, Sabrina and {Bhardwaj}, Mohit and {Boyce}, Michelle M. and {Boyle}, P.~J. and {Brar}, Charanjot and {Breitman}, Daniela and {Cassanelli}, Tomas and {Chawla}, Pragya and {Chen}, Tianyue and {Cliche}, J. -F. and {Cook}, Amanda and {Cubranic}, Davor and {Curtin}, Alice P. and {Deng}, Meiling and {Dobbs}, Matt and {Dong}, Fengqiu Adam and {Eadie}, Gwendolyn and {Fandino}, Mateus and {Fonseca}, Emmanuel and {Gaensler}, B.~M. and {Giri}, Utkarsh and {Good}, Deborah C. and {Halpern}, Mark and {Hill}, Alex S. and {Hinshaw}, Gary and {Josephy}, Alexander and {Kaczmarek}, Jane F. and {Kader}, Zarif and {Kania}, Joseph W. and {Kaspi}, Victoria M. and {Landecker}, T.~L. and {Lang}, Dustin and {Leung}, Calvin and {Li}, Dongzi and {Lin}, Hsiu-Hsien and {Masui}, Kiyoshi W. and {McKinven}, Ryan and {Mena-Parra}, Juan and {Merryfield}, Marcus and {Meyers}, Bradley W. and {Michilli}, Daniele and {Milutinovic}, Nikola and {Mirhosseini}, Arash and {M{\"u}nchmeyer}, Moritz and {Naidu}, Arun and {Newburgh}, Laura and {Ng}, Cherry and {Patel}, Chitrang and {Pen}, Ue-Li and {Petroff}, Emily and {Pinsonneault-Marotte}, Tristan and {Pleunis}, Ziggy and {Rafiei-Ravandi}, Masoud and {Rahman}, Mubdi and {Ransom}, Scott M. and {Renard}, Andre and {Sanghavi}, Pranav and {Scholz}, Paul and {Shaw}, J. Richard and {Shin}, Kaitlyn and {Siegel}, Seth R. and {Sikora}, Andrew E. and {Singh}, Saurabh and {Smith}, Kendrick M. and {Stairs}, Ingrid and {Tan}, Chia Min and {Tendulkar}, S.~P. and {Vanderlinde}, Keith and {Wang}, Haochen and {Wulf}, Dallas and {Zwaniga}, A.~V.},
        title = "{The First CHIME/FRB Fast Radio Burst Catalog}",
      journal = {\apjs},
         year = 2021,
        month = dec,
       volume = {257},
       number = {2},
          eid = {59},
        pages = {59},
          doi = {10.3847/1538-4365/ac33ab},
archivePrefix = {arXiv},
       eprint = {2106.04352},
 primaryClass = {astro-ph.HE},
       adsurl = {https://ui.adsabs.harvard.edu/abs/2021ApJS..257...59C}
}

@ARTICLE{2021ApJ...923L..17Z,
       author = {{Zhao}, Z.~Y. and {Wang}, F.~Y.},
        title = "{FRB 190520B Embedded in a Magnetar Wind Nebula and Supernova Remnant: A Luminous Persistent Radio Source, Decreasing Dispersion Measure, and Large Rotation Measure}",
      journal = {\apjl},
         year = 2021,
        month = dec,
       volume = {923},
       number = {1},
          eid = {L17},
        pages = {L17},
          doi = {10.3847/2041-8213/ac3f2f},
archivePrefix = {arXiv},
       eprint = {2112.00935},
 primaryClass = {astro-ph.HE},
       adsurl = {https://ui.adsabs.harvard.edu/abs/2021ApJ...923L..17Z}
}

@ARTICLE{2024ApJ...973..133D,
       author = {{Dong}, Y. and {Eftekhari}, T. and {Fong}, W. and {Bhandari}, S. and {Berger}, E. and {Ould-Boukattine}, O.~S. and {Hessels}, J.~W.~T. and {Sridhar}, N. and {Reines}, A. and {Margalit}, B. and {Darling}, J. and {Gordon}, A.~C. and {Greene}, J.~E. and {Kilpatrick}, C.~D. and {Marcote}, B. and {Metzger}, B.~D. and {Nimmo}, K. and {Nugent}, A.~E. and {Paragi}, Z. and {Williams}, P.~K.~G.},
        title = "{A Radio Study of Persistent Radio Sources in Nearby Dwarf Galaxies: Implications for Fast Radio Bursts}",
      journal = {\apj},
         year = 2024,
        month = oct,
       volume = {973},
       number = {2},
          eid = {133},
        pages = {133},
          doi = {10.3847/1538-4357/ad6568},
archivePrefix = {arXiv},
       eprint = {2405.00784},
 primaryClass = {astro-ph.HE},
       adsurl = {https://ui.adsabs.harvard.edu/abs/2024ApJ...973..133D}
}

@Article{         harris2020array,
 title         = {Array programming with {NumPy}},
 author        = {Charles R. Harris and K. Jarrod Millman and St{\'{e}}fan J.
                 van der Walt and Ralf Gommers and Pauli Virtanen and David
                 Cournapeau and Eric Wieser and Julian Taylor and Sebastian
                 Berg and Nathaniel J. Smith and Robert Kern and Matti Picus
                 and Stephan Hoyer and Marten H. van Kerkwijk and Matthew
                 Brett and Allan Haldane and Jaime Fern{\'{a}}ndez del
                 R{\'{i}}o and Mark Wiebe and Pearu Peterson and Pierre
                 G{\'{e}}rard-Marchant and Kevin Sheppard and Tyler Reddy and
                 Warren Weckesser and Hameer Abbasi and Christoph Gohlke and
                 Travis E. Oliphant},
 year          = {2020},
 month         = sep,
 journal       = {Nature},
 volume        = {585},
 number        = {7825},
 pages         = {357--362},
 doi           = {10.1038/s41586-020-2649-2},
 publisher     = {Springer Science and Business Media {LLC}},
 url           = {https://doi.org/10.1038/s41586-020-2649-2}
}

@ARTICLE{2020SciPy-NMeth,
  author  = {Virtanen, Pauli and Gommers, Ralf and Oliphant, Travis E. and
            Haberland, Matt and Reddy, Tyler and Cournapeau, David and
            Burovski, Evgeni and Peterson, Pearu and Weckesser, Warren and
            Bright, Jonathan and {van der Walt}, St{\'e}fan J. and
            Brett, Matthew and Wilson, Joshua and Millman, K. Jarrod and
            Mayorov, Nikolay and Nelson, Andrew R. J. and Jones, Eric and
            Kern, Robert and Larson, Eric and Carey, C J and
            Polat, {\.I}lhan and Feng, Yu and Moore, Eric W. and
            {VanderPlas}, Jake and Laxalde, Denis and Perktold, Josef and
            Cimrman, Robert and Henriksen, Ian and Quintero, E. A. and
            Harris, Charles R. and Archibald, Anne M. and
            Ribeiro, Ant{\^o}nio H. and Pedregosa, Fabian and
            {van Mulbregt}, Paul and {SciPy 1.0 Contributors}},
  title   = {{{SciPy} 1.0: Fundamental Algorithms for Scientific
            Computing in Python}},
  journal = {Nature Methods},
  year    = {2020},
  volume  = {17},
  pages   = {261--272},
  adsurl  = {https://rdcu.be/b08Wh},
  doi     = {10.1038/s41592-019-0686-2},
}

@Article{Hunter:2007,
  Author    = {Hunter, J. D.},
  Title     = {Matplotlib: A 2D graphics environment},
  Journal   = {Computing in Science \& Engineering},
  Volume    = {9},
  Number    = {3},
  Pages     = {90--95},
  publisher = {IEEE COMPUTER SOC},
  doi       = {10.1109/MCSE.2007.55},
  year      = 2007
}

@ARTICLE{astropy:2013,
       author = {{Astropy Collaboration} and {Robitaille}, Thomas P. and {Tollerud}, Erik J. and {Greenfield}, Perry and {Droettboom}, Michael and {Bray}, Erik and {Aldcroft}, Tom and {Davis}, Matt and {Ginsburg}, Adam and {Price-Whelan}, Adrian M. and {Kerzendorf}, Wolfgang E. and {Conley}, Alexander and {Crighton}, Neil and {Barbary}, Kyle and {Muna}, Demitri and {Ferguson}, Henry and {Grollier}, Fr{\'e}d{\'e}ric and {Parikh}, Madhura M. and {Nair}, Prasanth H. and {Unther}, Hans M. and {Deil}, Christoph and {Woillez}, Julien and {Conseil}, Simon and {Kramer}, Roban and {Turner}, James E.~H. and {Singer}, Leo and {Fox}, Ryan and {Weaver}, Benjamin A. and {Zabalza}, Victor and {Edwards}, Zachary I. and {Azalee Bostroem}, K. and {Burke}, D.~J. and {Casey}, Andrew R. and {Crawford}, Steven M. and {Dencheva}, Nadia and {Ely}, Justin and {Jenness}, Tim and {Labrie}, Kathleen and {Lim}, Pey Lian and {Pierfederici}, Francesco and {Pontzen}, Andrew and {Ptak}, Andy and {Refsdal}, Brian and {Servillat}, Mathieu and {Streicher}, Ole},
        title = "{Astropy: A community Python package for astronomy}",
      journal = {\aap},
         year = 2013,
        month = oct,
       volume = {558},
          eid = {A33},
        pages = {A33},
          doi = {10.1051/0004-6361/201322068},
archivePrefix = {arXiv},
       eprint = {1307.6212},
 primaryClass = {astro-ph.IM},
       adsurl = {https://ui.adsabs.harvard.edu/abs/2013A&A...558A..33A}
}

@ARTICLE{astropy:2018,
       author = {{Astropy Collaboration} and {Price-Whelan}, A.~M. and {Sip{\H{o}}cz}, B.~M. and {G{\"u}nther}, H.~M. and {Lim}, P.~L. and {Crawford}, S.~M. and {Conseil}, S. and {Shupe}, D.~L. and {Craig}, M.~W. and {Dencheva}, N. and {Ginsburg}, A. and {VanderPlas}, J.~T. and {Bradley}, L.~D. and {P{\'e}rez-Su{\'a}rez}, D. and {de Val-Borro}, M. and {Aldcroft}, T.~L. and {Cruz}, K.~L. and {Robitaille}, T.~P. and {Tollerud}, E.~J. and {Ardelean}, C. and {Babej}, T. and {Bach}, Y.~P. and {Bachetti}, M. and {Bakanov}, A.~V. and {Bamford}, S.~P. and {Barentsen}, G. and {Barmby}, P. and {Baumbach}, A. and {Berry}, K.~L. and {Biscani}, F. and {Boquien}, M. and {Bostroem}, K.~A. and {Bouma}, L.~G. and {Brammer}, G.~B. and {Bray}, E.~M. and {Breytenbach}, H. and {Buddelmeijer}, H. and {Burke}, D.~J. and {Calderone}, G. and {Cano Rodr{\'\i}guez}, J.~L. and {Cara}, M. and {Cardoso}, J.~V.~M. and {Cheedella}, S. and {Copin}, Y. and {Corrales}, L. and {Crichton}, D. and {D'Avella}, D. and {Deil}, C. and {Depagne}, {\'E}. and {Dietrich}, J.~P. and {Donath}, A. and {Droettboom}, M. and {Earl}, N. and {Erben}, T. and {Fabbro}, S. and {Ferreira}, L.~A. and {Finethy}, T. and {Fox}, R.~T. and {Garrison}, L.~H. and {Gibbons}, S.~L.~J. and {Goldstein}, D.~A. and {Gommers}, R. and {Greco}, J.~P. and {Greenfield}, P. and {Groener}, A.~M. and {Grollier}, F. and {Hagen}, A. and {Hirst}, P. and {Homeier}, D. and {Horton}, A.~J. and {Hosseinzadeh}, G. and {Hu}, L. and {Hunkeler}, J.~S. and {Ivezi{\'c}}, {\v{Z}}. and {Jain}, A. and {Jenness}, T. and {Kanarek}, G. and {Kendrew}, S. and {Kern}, N.~S. and {Kerzendorf}, W.~E. and {Khvalko}, A. and {King}, J. and {Kirkby}, D. and {Kulkarni}, A.~M. and {Kumar}, A. and {Lee}, A. and {Lenz}, D. and {Littlefair}, S.~P. and {Ma}, Z. and {Macleod}, D.~M. and {Mastropietro}, M. and {McCully}, C. and {Montagnac}, S. and {Morris}, B.~M. and {Mueller}, M. and {Mumford}, S.~J. and {Muna}, D. and {Murphy}, N.~A. and {Nelson}, S. and {Nguyen}, G.~H. and {Ninan}, J.~P. and {N{\"o}the}, M. and {Ogaz}, S. and {Oh}, S. and {Parejko}, J.~K. and {Parley}, N. and {Pascual}, S. and {Patil}, R. and {Patil}, A.~A. and {Plunkett}, A.~L. and {Prochaska}, J.~X. and {Rastogi}, T. and {Reddy Janga}, V. and {Sabater}, J. and {Sakurikar}, P. and {Seifert}, M. and {Sherbert}, L.~E. and {Sherwood-Taylor}, H. and {Shih}, A.~Y. and {Sick}, J. and {Silbiger}, M.~T. and {Singanamalla}, S. and {Singer}, L.~P. and {Sladen}, P.~H. and {Sooley}, K.~A. and {Sornarajah}, S. and {Streicher}, O. and {Teuben}, P. and {Thomas}, S.~W. and {Tremblay}, G.~R. and {Turner}, J.~E.~H. and {Terr{\'o}n}, V. and {van Kerkwijk}, M.~H. and {de la Vega}, A. and {Watkins}, L.~L. and {Weaver}, B.~A. and {Whitmore}, J.~B. and {Woillez}, J. and {Zabalza}, V. and {Astropy Contributors}},
        title = "{The Astropy Project: Building an Open-science Project and Status of the v2.0 Core Package}",
      journal = {\aj},
         year = 2018,
        month = sep,
       volume = {156},
       number = {3},
          eid = {123},
        pages = {123},
          doi = {10.3847/1538-3881/aabc4f},
archivePrefix = {arXiv},
       eprint = {1801.02634},
 primaryClass = {astro-ph.IM},
       adsurl = {https://ui.adsabs.harvard.edu/abs/2018AJ....156..123A}
}

@ARTICLE{astropy:2022,
       author = {{Astropy Collaboration} and {Price-Whelan}, Adrian M. and {Lim}, Pey Lian and {Earl}, Nicholas and {Starkman}, Nathaniel and {Bradley}, Larry and {Shupe}, David L. and {Patil}, Aarya A. and {Corrales}, Lia and {Brasseur}, C.~E. and {N{\"o}the}, Maximilian and {Donath}, Axel and {Tollerud}, Erik and {Morris}, Brett M. and {Ginsburg}, Adam and {Vaher}, Eero and {Weaver}, Benjamin A. and {Tocknell}, James and {Jamieson}, William and {van Kerkwijk}, Marten H. and {Robitaille}, Thomas P. and {Merry}, Bruce and {Bachetti}, Matteo and {G{\"u}nther}, H. Moritz and {Aldcroft}, Thomas L. and {Alvarado-Montes}, Jaime A. and {Archibald}, Anne M. and {B{\'o}di}, Attila and {Bapat}, Shreyas and {Barentsen}, Geert and {Baz{\'a}n}, Juanjo and {Biswas}, Manish and {Boquien}, M{\'e}d{\'e}ric and {Burke}, D.~J. and {Cara}, Daria and {Cara}, Mihai and {Conroy}, Kyle E. and {Conseil}, Simon and {Craig}, Matthew W. and {Cross}, Robert M. and {Cruz}, Kelle L. and {D'Eugenio}, Francesco and {Dencheva}, Nadia and {Devillepoix}, Hadrien A.~R. and {Dietrich}, J{\"o}rg P. and {Eigenbrot}, Arthur Davis and {Erben}, Thomas and {Ferreira}, Leonardo and {Foreman-Mackey}, Daniel and {Fox}, Ryan and {Freij}, Nabil and {Garg}, Suyog and {Geda}, Robel and {Glattly}, Lauren and {Gondhalekar}, Yash and {Gordon}, Karl D. and {Grant}, David and {Greenfield}, Perry and {Groener}, Austen M. and {Guest}, Steve and {Gurovich}, Sebastian and {Handberg}, Rasmus and {Hart}, Akeem and {Hatfield-Dodds}, Zac and {Homeier}, Derek and {Hosseinzadeh}, Griffin and {Jenness}, Tim and {Jones}, Craig K. and {Joseph}, Prajwel and {Kalmbach}, J. Bryce and {Karamehmetoglu}, Emir and {Ka{\l}uszy{\'n}ski}, Miko{\l}aj and {Kelley}, Michael S.~P. and {Kern}, Nicholas and {Kerzendorf}, Wolfgang E. and {Koch}, Eric W. and {Kulumani}, Shankar and {Lee}, Antony and {Ly}, Chun and {Ma}, Zhiyuan and {MacBride}, Conor and {Maljaars}, Jakob M. and {Muna}, Demitri and {Murphy}, N.~A. and {Norman}, Henrik and {O'Steen}, Richard and {Oman}, Kyle A. and {Pacifici}, Camilla and {Pascual}, Sergio and {Pascual-Granado}, J. and {Patil}, Rohit R. and {Perren}, Gabriel I. and {Pickering}, Timothy E. and {Rastogi}, Tanuj and {Roulston}, Benjamin R. and {Ryan}, Daniel F. and {Rykoff}, Eli S. and {Sabater}, Jose and {Sakurikar}, Parikshit and {Salgado}, Jes{\'u}s and {Sanghi}, Aniket and {Saunders}, Nicholas and {Savchenko}, Volodymyr and {Schwardt}, Ludwig and {Seifert-Eckert}, Michael and {Shih}, Albert Y. and {Jain}, Anany Shrey and {Shukla}, Gyanendra and {Sick}, Jonathan and {Simpson}, Chris and {Singanamalla}, Sudheesh and {Singer}, Leo P. and {Singhal}, Jaladh and {Sinha}, Manodeep and {Sip{\H{o}}cz}, Brigitta M. and {Spitler}, Lee R. and {Stansby}, David and {Streicher}, Ole and {{\v{S}}umak}, Jani and {Swinbank}, John D. and {Taranu}, Dan S. and {Tewary}, Nikita and {Tremblay}, Grant R. and {de Val-Borro}, Miguel and {Van Kooten}, Samuel J. and {Vasovi{\'c}}, Zlatan and {Verma}, Shresth and {de Miranda Cardoso}, Jos{\'e} Vin{\'\i}cius and {Williams}, Peter K.~G. and {Wilson}, Tom J. and {Winkel}, Benjamin and {Wood-Vasey}, W.~M. and {Xue}, Rui and {Yoachim}, Peter and {Zhang}, Chen and {Zonca}, Andrea and {Astropy Project Contributors}},
        title = "{The Astropy Project: Sustaining and Growing a Community-oriented Open-source Project and the Latest Major Release (v5.0) of the Core Package}",
      journal = {\apj},
         year = 2022,
        month = aug,
       volume = {935},
       number = {2},
          eid = {167},
        pages = {167},
          doi = {10.3847/1538-4357/ac7c74},
archivePrefix = {arXiv},
       eprint = {2206.14220},
 primaryClass = {astro-ph.IM},
       adsurl = {https://ui.adsabs.harvard.edu/abs/2022ApJ...935..167A}
}

@ARTICLE{2024Natur.632.1014B,
       author = {{Bruni}, Gabriele and {Piro}, Luigi and {Yang}, Yuan-Pei and {Quai}, Salvatore and {Zhang}, Bing and {Palazzi}, Eliana and {Nicastro}, Luciano and {Feruglio}, Chiara and {Tripodi}, Roberta and {O'Connor}, Brendan and {Gardini}, Angela and {Savaglio}, Sandra and {Rossi}, Andrea and {Nicuesa Guelbenzu}, Ana M. and {Paladino}, Rosita},
        title = "{A nebular origin for the persistent radio emission of fast radio bursts}",
      journal = {\nat},
         year = 2024,
        month = aug,
       volume = {632},
       number = {8027},
        pages = {1014-1016},
          doi = {10.1038/s41586-024-07782-6},
archivePrefix = {arXiv},
       eprint = {2312.15296},
 primaryClass = {astro-ph.HE},
       adsurl = {https://ui.adsabs.harvard.edu/abs/2024Natur.632.1014B}
}

@ARTICLE{2018IMMag..19..112L,
       author = {{Li}, Di and {Wang}, Pei and {Qian}, Lei and {Krco}, Marko and {Jiang}, Peng and {Yue}, Youling and {Jin}, Chenjin and {Zhu}, Yan and {Pan}, Zhichen and {Nan}, Rendong and {Dunning}, Alex},
        title = "{FAST in Space: Considerations for a Multibeam, Multipurpose Survey Using China's 500-m Aperture Spherical Radio Telescope (FAST)}",
      journal = {IEEE Microwave Magazine},
         year = 2018,
        month = apr,
       volume = {19},
       number = {3},
        pages = {112-119},
          doi = {10.1109/MMM.2018.2802178},
archivePrefix = {arXiv},
       eprint = {1802.03709},
 primaryClass = {astro-ph.IM},
       adsurl = {https://ui.adsabs.harvard.edu/abs/2018IMMag..19..112L}
}

@ARTICLE{2025ApJ...982..203C,
       author = {{Chen}, Xiang-Lei and {Tsai}, Chao-Wei and {Stern}, Daniel and {Bochenek}, Christopher D. and {Chatterjee}, Shami and {Law}, Casey and {Li}, Di and {Niu}, Chen-hui and {Niino}, Yuu and {Feng}, Yi and {Wang}, Pei and {Assef}, Roberto J. and {Li}, Guo-dong and {Lake}, Sean E. and {Luo}, Gan and {Liao}, Mai},
        title = "{The Host Galaxy of FRB 20190520B and Its Unique Ionized Gas Distribution}",
      journal = {\apj},
         year = 2025,
        month = apr,
       volume = {982},
       number = {2},
          eid = {203},
        pages = {203},
          doi = {10.3847/1538-4357/adb84d},
archivePrefix = {arXiv},
       eprint = {2503.01740},
 primaryClass = {astro-ph.GA},
       adsurl = {https://ui.adsabs.harvard.edu/abs/2025ApJ...982..203C}
}

@ARTICLE{2023ApJ...954L...7L,
       author = {{Lee}, Khee-Gan and {Khrykin}, Ilya S. and {Simha}, Sunil and {Ata}, Metin and {Huang}, Yuxin and {Prochaska}, J. Xavier and {Tejos}, Nicolas and {Cooke}, Jeff and {Nagamine}, Kentaro and {Zhang}, Jielai},
        title = "{The FRB 20190520B Sight Line Intersects Foreground Galaxy Clusters}",
      journal = {\apjl},
         year = 2023,
        month = sep,
       volume = {954},
       number = {1},
          eid = {L7},
        pages = {L7},
          doi = {10.3847/2041-8213/acefb5},
archivePrefix = {arXiv},
       eprint = {2306.05403},
 primaryClass = {astro-ph.GA},
       adsurl = {https://ui.adsabs.harvard.edu/abs/2023ApJ...954L...7L}
}

@ARTICLE{2023Univ....9..330X,
       author = {{Xu}, Jiaying and {Feng}, Yi and {Li}, Di and {Wang}, Pei and {Zhang}, Yongkun and {Xie}, Jintao and {Chen}, Huaxi and {Wang}, Han and {Kang}, Zhixuan and {Hu}, Jingjing and {Zheng}, Yun and {Tsai}, Chao-Wei and {Chen}, Xianglei and {Zhou}, Dengke},
        title = "{Blinkverse: A Database of Fast Radio Bursts}",
      journal = {Universe},
         year = 2023,
        month = jul,
       volume = {9},
       number = {7},
          eid = {330},
        pages = {330},
          doi = {10.3390/universe9070330},
archivePrefix = {arXiv},
       eprint = {2308.00336},
 primaryClass = {astro-ph.HE},
       adsurl = {https://ui.adsabs.harvard.edu/abs/2023Univ....9..330X}
}

@ARTICLE{2024ApJ...976..165Y,
       author = {{Yang}, Ai Yuan and {Feng}, Yi and {Tsai}, Chao-Wei and {Li}, Di and {Shi}, Hui and {Wang}, Pei and {Yang}, Yuan-Pei and {Zhang}, Yong-Kun and {Niu}, Chen-Hui and {Yao}, Ju-Mei and {Cui}, Yu-Zhu and {Su}, Ren-Zhi and {Li}, Xiao-Feng and {Zhang}, Jun-Shuo and {Zhu}, Yu-Hao and {Cotton}, W.~D.},
        title = "{The Variability of Persistent Radio Sources of Fast Radio Bursts}",
      journal = {\apj},
         year = 2024,
        month = dec,
       volume = {976},
       number = {2},
          eid = {165},
        pages = {165},
          doi = {10.3847/1538-4357/ad7d02},
archivePrefix = {arXiv},
       eprint = {2409.13170},
 primaryClass = {astro-ph.HE},
       adsurl = {https://ui.adsabs.harvard.edu/abs/2024ApJ...976..165Y}
}

@article{Hussain2019pyMannKendall,
	journal = {Journal of Open Source Software},
	doi = {10.21105/joss.01556},
	issn = {2475-9066},
	number = {39},
	publisher = {The Open Journal},
	title = {pyMannKendall: a python package for non parametric Mann Kendall family of trend tests.},
	url = {http://dx.doi.org/10.21105/joss.01556},
	volume = {4},
	author = {Hussain, Md. and Mahmud, Ishtiak},
	pages = {1556},
	date = {2019-07-25},
	year = {2019},
	month = {7},
	day = {25},
}

@ARTICLE{2016MNRAS.461.1498M,
       author = {{Murase}, Kohta and {Kashiyama}, Kazumi and {M{\'e}sz{\'a}ros}, Peter},
        title = "{A burst in a wind bubble and the impact on baryonic ejecta: high-energy gamma-ray flashes and afterglows from fast radio bursts and pulsar-driven supernova remnants}",
      journal = {\mnras},
         year = 2016,
        month = sep,
       volume = {461},
       number = {2},
        pages = {1498-1511},
          doi = {10.1093/mnras/stw1328},
archivePrefix = {arXiv},
       eprint = {1603.08875},
 primaryClass = {astro-ph.HE},
       adsurl = {https://ui.adsabs.harvard.edu/abs/2016MNRAS.461.1498M}
}

@Article{universe11070206,
AUTHOR = {Xu, Jiaying and Tsai, Chao-Wei and Lake, Sean E. and Feng, Yi and Chen, Xiang-Lei and Li, Di and Wang, Han and Guo, Xuerong and Hu, Jingjing and Ge, Xiaodong},
TITLE = {Blinkverse 2.0: Updated Host Galaxies for Fast Radio Bursts},
JOURNAL = {Universe},
VOLUME = {11},
YEAR = {2025},
NUMBER = {7},
ARTICLE-NUMBER = {206},
URL = {https://www.mdpi.com/2218-1997/11/7/206},
ISSN = {2218-1997},
DOI = {10.3390/universe11070206}
}
\bibliographystyle{aasjournal}

\appendix 
\restartappendixnumbering

\section{Comparison of calibration of images }

\begin{figure*}[h]
    \centering
    \includegraphics[width=0.9\linewidth]{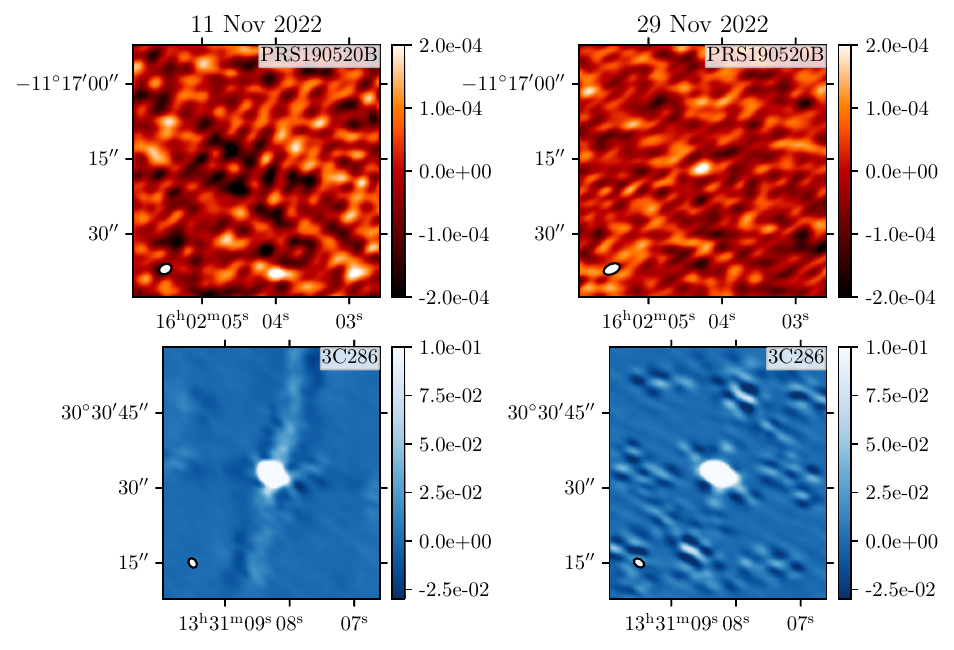}
    \caption{Comparison of the image of the target PRS (top row) and the calibrator 3C286 (bottom row) on 11 Nov 2022 (left column) and 29 Nov 2022 (right column). This shows the calibration errors in the 11 Nov 2022 data.}
    \label{fig:upper_limits}
\end{figure*}

This section contains a comparison of the image of the
target PRS along with the calibrator, 3C286 to demonstrate the
calibrations errors present in the 2022 November 11 data. This
is shown in Figure \ref{fig:upper_limits}.

\end{document}